\documentclass[12pt]{iopart}

\usepackage{graphicx}
\usepackage{color}
\usepackage{soul}
\usepackage{url}
\usepackage{bm}         
\usepackage{times}
\usepackage{dcolumn}
\usepackage{bm}
\usepackage{epsf}
\usepackage{amssymb}
\usepackage{booktabs}
\usepackage{tensor}
\usepackage{siunitx}
\usepackage[caption=false]{subfig}
\usepackage[%
  bookmarksopen=true,
  colorlinks=true,
  urlcolor=blue,
  linkcolor=blue,
  citecolor=blue
]{hyperref}

\newcommand{\sdetg}{\sqrt{\gamma}}
\newcommand{\pdv}[2]{\frac{\partial #1}{\partial #2}}
\newcommand{\intd}[1]{\mathrm{d}#1}
\newcommand{\Msun}{\ensuremath{\text{M}_\odot}}
\newcommand{\lmix}{\ensuremath{\ell_{\rm mix}}}
\newcommand{\ind}[1]{\indices{#1}}

\begin{document}
\title{Axisymmetric Hydrodynamics in Numerical Relativity Using a Multipatch Method}


\author{
Jerred Jesse$^{1}$,
Matthew D. Duez$^{1}$,
Francois Foucart$^{2}$,
Milad Haddadi$^1$,
Alexander L. Knight$^2$,
Courtney L. Cadenhead$^1$,
Francois H\'{e}bert$^{3}$,
Lawrence E. Kidder$^4$,
Harald P. Pfeiffer$^5$,
Mark A. Scheel$^3$
}

\address{$^1$ Department of Physics \& Astronomy,       Washington State University, Pullman, Washington 99164, USA}
\address{$^2$ Department of Physics \& Astronomy, University of New Hampshire, 9 Library Way, Durham NH 03824, USA}
\address{$^3$ TAPIR, Walter Burke Institute for Theoretical Physics, MC 350-17, California Institute of Technology, Pasadena, California 91125, USA}
\address{$^4$ Center for Radiophysics and Space Research, Cornell University, Ithaca, New York, 14853, USA}
\address{$^5$ Max-Planck-Institut fur Gravitationsphysik, Albert-Einstein-Institut, D-14476 Golm, Germany}

\begin{abstract}
We describe a method of implementing the axisymmetric evolution of general-relativistic hydrodynamics and magnetohydrodynamics through modification of a multipatch grid scheme. In order to ease the computational requirements required to evolve the post-merger phase of systems involving binary compact massive objects in numerical relativity, it is often beneficial to take advantage of these system's tendency to rapidly settle into states that are nearly axisymmetric, allowing for 2D evolution of secular timescales. We implement this scheme in the Spectral Einstein Code (SpEC) and show the results of application of this method to four test systems including viscosity, magnetic fields, and neutrino radiation transport. Our results show that this method can be used to quickly allow already existing 3D infrastructure that makes use of local coordinate system transformations to be made to run in axisymmetric 2D with the flexible grid creation capabilities of multipatch methods.  Our code tests include a simple model of a binary neutron star postmerger remnant, for which we confirm the formation of a massive torus which is a promising source of post-merger ejecta.
\end{abstract}

\maketitle

\section{Introduction} \label{sec:introduction}

The detection of the gravitational wave signal resulting from the merger binary black hole systems by the LIGO and VIRGO collaborations \cite{Abbott2016a,Abbott2016b,Abbott2016c,Abbott2017a,Abbott2017b,Abbott2017c,Abbott2019} along with the detection of simultaneous electromagnetic and gravitational wave signals from binary neutron star mergers \cite{Abbott2017d,Abbott2017e,Abbott2017f,Abbott2020}, and the corresponding need for theoretical predictions with which to compare them, has given renewed urgency to the goal of accurately modeling these systems throughout the merger process. For systems involving at least one neutron star, it is the post-merger state that is primarily responsible for the observable electromagnetic signals.  Modeling of these systems through numerical relativity simulations provides critical insight into the dependencies of the signals on binary parameters and nuclear physics.  Unfortunately, running these simulations in the high-resolution required to get accurate predictions can present large computational resource barriers in simulated time or size scales. However, the post-merger environment has a useful property: by taking advantage of these systems' tendency to approach an axisymmetric state, we can ease the computational resources required to simulate these systems over extended scales of both time and space. Although the dynamical timescales of remnant neutron stars and accretion disks, of the order $\sim$\si{ms} at most, are reasonably accessible to 3D simulations, secular effects that drive the subsequent evolution can operate on much longer timescales.  Particularly important are angular momentum transport effects that can act on a wide range of timescales of up to hundreds of milliseconds \cite{Shibata2017b, Hotokezaka2013}, and neutrino cooling effects that operate on timescales of up to several seconds \cite{Paschalidis2012}.

The use of axisymmetry in numerical relativity simulations has been explored by several groups. This typically involves evolving Einstein's equations using the cartoon method \cite{Alcubierre2001} while evolving hydrodynamics by writing the relevant equations in a cylindrical coordinate system \cite{Duez2004,Duez2006,Shibata2003}. The cartoon method does involve some loss of accuracy due to interpolations required in the method, and some effort has been made to avoid these \cite{Pretorius2005}. Additionally, evolution problems due to the coordinate singularities that arise from the use of polar coordinate systems have been avoided by use of a reference metric \cite{Baumgarte2013,Montero2014,Baumgarte2015}.  Methods also exist that help with issues of spatial resolution on large scales, such as adaptive mesh refinement~\cite{Pretorius2005,Anderson2008}, which is able to concentrate resolution where it is most needed, while in most cases still building the grid from Cartesian domains. In multipatch methods~\cite{Koldoba:2002kx,Schnetter2006,Reisswig:2006nt,Zink:2007xn,Fragile:2008ca,Nouri2018}, one introduces coordinate patches, each with its own local coordinate system in which it takes a simple shape (e.g. a Cartesian block), but which can be deformed in the global coordinate system and fit together into a grid to match the geometry of the problem. A number of methods used in numerical relativity not usually called ``multipatch'' have local coordinate systems and therefore fit into this general category, including the multidomain pseudospectral sector of the Spectral Einstein Code~\cite{SpEC2020} and the multielement discontinuous Galerkin methods~\cite{Hebert2018,Kidder2017} which many hope will form the basis of the next generation of numerical relativity codes.

In this paper, we describe a method of implementing the axisymmetric evolution of the general-relativistic equations of ideal radiation hydrodynamics and magnetohydrodynamics through modification of a multipatch grid scheme, applicable to any method using the local patch coordinates framework, which we implement in the Spectral Einstein Code (SpEC)~\cite{SpEC2020}.  While other codes for carrying out axisymmetric relativistic hydrodynamics evolutions exist, there are several notable new features of our methods and results.  First, using the multipatch framework, we automate the conversion to an axisymmetry-friendly coordinate system, which we demonstrate by using the same recipe to make working axisymmetric 2D versions of our relativistic 3D hydrodynamics, magnetohydrodynamics, neutrino transport, and shear viscosity codes.  We point out that our neutrino transport method, which evolves number density as well as energy density, is somewhat different from other grey M1 schemes, so this is the first time this particular system has been converted to 2D axisymmetry.  Second, we  inherit the flexibility of multipatch methods to construct grids from patches of different shapes to optimally match the geometry of a problem.  As a demonstration of this, we present the 2D evolution of a viscous differentially rotating star using a combination of square patches for the stellar interior and circular wedges for the outflow zone.  As well as serving as a code test, this viscous rotating star system is of great astrophysical interest because it is a reasonable model of the remnant of a binary neutron star merger.  We evolve it using a different subgrid momentum transport model than has been applied to it in prior work~\cite{Shibata2017a}, providing an important qualitative check on the previous results.  Finally, we present several minor enhancements of the SpEC-Hydro code, including a generalization of our auxiliary entropy evolution~\cite{Nouri2018} to the thermal Gamma-law class of equations of state, an altered form of the divergence cleaning magnetohydrodynamics equations, and an altered treatment of neutrino fluxes in the optically thick limit.

This paper is organized as follows.  In Sec. \ref{sec:formulation}, we describe the evolution equations for our (magneto)hydrodynamic variables and the application of our axisymmetry method to them. In Sec. \ref{sec:tests} several tests of this axisymmetry method are presented: a stationary TOV star, a viscous differentially rotating star, a magnetized accretion disk, and neutrino radiation in a spherically symmetric supernova collapse profile, each showing good agreement with 3D results or previous axisymmetric simulations. Concluding remarks are given in Sec. \ref{sec:conclusion}, where we summarize our results and discuss future plans.

\section{Formulation} \label{sec:formulation}

\subsection{Evolution Equations}
We use SpEC to evolve Einstein's equations and the general relativistic equations of ideal radiation (magneto)hydrodynamics. SpEC evolves Einstein's equations and the general relativistic hydrodynamics equations on two separate computational grids. A multidomain grid of colocation points is used to evolve Einstein's equations pseudospectrally in a generalized harmonic formulation \cite{Lindblom2006} while the general relativistic (magneto)hydrodynamics equations in conservative form are evolved on a finite difference grid. The finite difference grid uses an HLL approximate Riemann solver \cite{Harten1983}. Reconstruction of values at cell faces from their cell-average values is done using a high-order shock capturing method, a fifth-order WENO scheme \cite{Liu1994, Jiang1996}. Time evolution is performed using a third-order Runge-Kutta algorithm with an adaptive time-stepper which compares integration errors to specified absolute and relative thresholds, lowering the timestep if errors exceed thresholds, increasing the timestep if they are all below thresholds.  At the end of each time step any necessary source term information is then communicated between the two grids, using a third-order accurate spatial interpolation scheme \cite{Duez2008}.  Both grids are forced to evolve with the same timestep, determined by the grid which requires the shorter step.  Interpolation between grids is carried out once per step, and second-order interpolation in time is used for source terms during Runge-Kutta substeps.

The following sections make use of the 3+1 decomposition of the spacetime metric
\begin{eqnarray}
\label{eq:metric}
ds^2 &=& g\ind{_\alpha_\beta} dx^\alpha dx^\beta \\
   &=& -\alpha^2 dt^2 + \gamma\ind{_i_j} \left(dx^i + \beta^i dt \right) \left(dx^j + \beta^j dt \right),
\end{eqnarray}
where $\alpha$ is the lapse, $\beta^i$ the shift, and $\gamma\ind{_i_j}$ is the three-metric on a spacelike hypersurface of constant coordinate $t$. The three-metric is the projection onto spatial hypersurfaces of the four-metric:
\begin{equation}
\label{eq:three-metric}
\gamma\ind{_i_j} = g\ind{_i_j} + n_i n_j,
\end{equation}
where $n_\mu = (-\alpha, 0, 0, 0)$ is the unit normal to the $t = $ constant hypersurface. Additionally, we use the units with
$G = c = 1$ throughout.

\subsubsection{Fluid}
We begin by treating our fluid as a perfect fluid with the stress-energy tensor
\begin{equation}
\label{eq:fluid-se-tensor}
T\ind{_\mu_\nu} = \rho_0 h u_\mu u_\nu + P g\ind{_\mu_\nu},
\end{equation}
where $\rho_0$ is the baryon density, $h = 1+P/\rho_0+\epsilon$ is the specific
enthalpy, $P$ is the pressure, $u_\mu$ the four-velocity, and $\epsilon$ the specific internal energy.

The general relativistic hydrodynamics equations are evolved using the conservative variables
\begin{eqnarray}
\label{eq:fluid-density}
\rho_* &=& - \sdetg n_\mu n^\nu \rho_0 = \rho_0 W \sdetg, \\
\label{eq:fluid-energy}
\tau &=& \sdetg n_\mu n_\nu T\ind{^\mu^\nu} - \rho_* = \rho_* \left(hW - 1 \right) - P\sdetg, \\
\label{eq:fluid-momentum}
S_i &=& - \sdetg n_{\mu} T\ind{_i^\mu} = \rho_* h u_i,
\end{eqnarray}
where $W = \sqrt{1 + \gamma\ind{^i^j} u_i u_j}$ is the Lorentz factor and $\gamma$ is the determinant of $\gamma\ind{_i_j}$.
Using conservation of energy and momentum, $\nabla_\nu T\ind{^\mu^\nu} = 0$, and baryon number conservation, $\nabla_\mu \left(\rho_0 u^\mu \right) = 0 $, we get the evolution equations for the conservative variables:
\begin{eqnarray}
\label{eq:fluid-density-evolution}
\partial_t \rho_* &+& \partial_j \left(\rho_* v_T^j \right) = 0, \\
\label{eq:fluid-energy-evolution}
\partial_t \tau &+& \partial_j \left(\alpha^2\sdetg T\ind{^0^i} - \rho_* {v_T}^i \right) = - \alpha \sdetg T\ind{^\mu^\nu} \nabla_\mu n_\nu, \\
\label{eq:fluid-momentum-evolution}
\partial_t S_i &+& \partial_j \left(\alpha\sdetg T\ind{_i^j}\right) = \frac{1}{2} \alpha \sdetg T\ind{^\mu^\nu} \partial_i g\ind{_\mu_\nu},
\end{eqnarray}
where the Eulerian velocity $v^i$ is related to the fluid transport velocity ${v_T}^i$ by ${v_T}^i = \alpha v^i - \beta^i$. Additionally, for simulations involving nuclear matter and neutrinos, we evolve the electron fraction of the fluid, $Y_e$,
\begin{equation}
\label{eq:fluid-composition-evolution}
\partial_t \left(\rho_* Y_e\right) + \partial_j \left(\rho_* Y_e v_T^j \right) = 0.
\end{equation}

To close these equations we must also supply an equation of state for the pressure and enthalpy: $P = P(\rho_*, T, Y_e)$ and $h = h(\rho_*, T, Y_e)$.

\subsubsection{Magnetic Fields}
To handle magnetic fields, we begin by adding the electromagnetic contribution, ${T_{\rm EM}}\ind{^\mu^\nu}$, to the fluid stress-energy tensor, where
\begin{equation}
\label{eq:mag-se-tensor}
{T_{\rm EM}}\ind{^\mu^\nu} = F\ind{^\mu^\alpha} F\ind{^\nu_\alpha} - \frac{1}{4} F\ind{_\alpha_\beta} F\ind{^\alpha^\beta} g\ind{^\mu^\nu},
\end{equation}
and $F\ind{^\mu^\nu}$ is the Faraday tensor. We treat the fluid as a perfect conductor, $F\ind{^\mu^\nu} u_\nu = 0$, which gives an electric field which can be computed from velocity and magnetic field.

We use two different methods for evolving the magnetic field, as described in \cite{Nouri2018}. The first method evolves the magnetic vector potential $A_i$ and scalar potential $\Phi$.  In the generalized Lorentz gauge~\cite{Farris:2012ux}, the most robust gauge choice we have explored, the evolution equations are
\begin{eqnarray}
\label{eq:mag-vector-potential-evolution}
\partial_t A_i + \partial_i \left(\alpha \Phi - \beta^j A_j \right) &=& \epsilon\ind{_i_j_k} v^j B^k, \\
\label{eq:mag-scalar-potential-evolution}
\partial_t \left(\sdetg \Phi \right) + \partial_j \left(\alpha \sdetg A^j - \sdetg \beta^j \Phi \right) &=& - \xi \alpha \sdetg \Phi,
\end{eqnarray}
where $\xi$ is a specifiable constant of the order of the mass of the system.

The second method evolves the magnetic field using a covariant hyperbolic divergence cleaning method \cite{Liebling2010,Penner2011,Mosta2013} in which an auxiliary scalar evolution variable $\Psi$ is introduced in order to propagate and damp monopole formation.
In this method, the induction equation takes the form
\begin{eqnarray}
\label{eq:mag-div-clean-b-evolution}
\partial_t \tilde{B}^i - \partial_i \left(v^j \tilde{B}^i - v^i \tilde{B}^j \right) &=& \alpha \gamma\ind{^i^j} \partial_j \tilde{\Psi} + \beta^i \partial_j \tilde{B}^j, \\
\label{eq:mag-div-clean-psi-evolution}
\partial_t \tilde{\Psi} + \partial_i \left(\alpha \tilde{B}^i - \beta^i \tilde{\Psi} \right) &=& \tilde{B}^i \partial_i \alpha - \alpha \left(K\indices{^i_i} + \lambda \right) \tilde{\Psi},
\end{eqnarray}
where $\tilde{B}^i = \sdetg B^i$, $\tilde{\Psi} = \sdetg \Psi$, $K\indices{^i_i}$ is the trace of the extrinsic curvature, and $\lambda$ is a specifiable damping constant.  Previously~\cite{Nouri2018}, we had used $\Psi$ rather than $\tilde{\Psi}$ as an evolution variable, but we find the new choice to be slightly more robust near excision inner boundaries.

\subsubsection{Neutrinos}

Neutrino evolution is handled using the gray two-moment scheme as described in \cite{Foucart2015,Foucart2016}. This method provides evolution of neutrino average energy densities, flux densities, and number densities. We define three neutrino species that we evolve: electron neutrinos $\nu_e$, electron antineutrinos $\bar{\nu}_e$, and the heavy lepton neutrinos $\nu_x$. The heavy lepton neutrino species groups together the four heavy lepton neutrinos and antineutrinos: $\nu_\mu$, $\bar{\nu}_\mu$, $\nu_\tau$, and $\bar{\nu}_\tau$.

We can describe each of our three species of neutrinos $\nu_i$ using each species' distribution function $f_\nu\left(x^\mu,p^\mu\right)$, where $x^\mu=\left(t,x^i\right)$ gives the time and position of the neutrinos and $p^\mu$ is the 4-momentum of the neutrinos. $f_\nu$ evolves in phase space according to the Boltzmann transport equation:
\begin{equation}
\label{eq:m1nu-distribution}
p^\alpha \left[\pdv{f_{\left(\nu\right)}}{x^\alpha} - \Gamma\ind{^\beta_\alpha_\gamma}p^\gamma \pdv{f_{\left(\nu\right)}}{p^\beta}\right] = C\left[f_{\left(\nu\right)}\right],
\end{equation}
where the term $C\left[f_{\left(\nu\right)}\right]$ includes all collisional processes (emissions, absorptions, and scatterings).

We simplify the radiation evolution by taking the gray approximation (integrating over the neutrino spectrum) and evolving the lowest two moments of the distribution functions of each neutrino species, truncating the moment expansion by imposing the Minerbo closure~\cite{Minerbo1978}.  Our evolved quantities are
projections of the stress-energy tensor of the neutrino radiation, ${T_{\rm rad}}\ind{^\mu^\nu}$. The decomposition of ${T_{\rm rad}}\ind{^\mu^\nu}$ in the fluid frame is
\begin{equation}
\label{eq:m1nu-se-tensor}
{T_{\rm rad}}\ind{^\mu^\nu} = J u^\mu u^\nu + H^\mu u^\nu + H^\nu u^\mu + S\ind{^\mu^\nu}
\end{equation}
with $H^\mu u_\mu = S\ind{^\mu^\nu} u_\mu = 0$. The energy density $J$, flux density $H^\mu$, and stress density $S\ind{^\mu^\nu}$ of the neutrino radiation as observed in the frame comoving with the fluid are related to the distribution functions by
\begin{eqnarray}
\label{eq:m1nu-j}
J &=& \int_{0}^{\infty} \intd{\nu}\, \nu^3 \int \intd{\Omega}\, f_{\left(\nu\right)}\left(x^\alpha,\nu,\Omega\right), \\
\label{eq:m1nu-h}
H^\mu &=& \int_{0}^{\infty} \intd{\nu}\, \nu^3 \int \intd{\Omega}\, f_{\left(\nu\right)}\left(x^\alpha,\nu,\Omega\right)l^\mu, \\
\label{eq:m1nu-s}
S\ind{^\mu^\nu} &=& \int_{0}^{\infty} \intd{\nu}\, \nu^3 \int \intd{\Omega}\, f_{\left(\nu\right)}\left(x^\alpha,\nu,\Omega\right)l^\mu l^\nu,
\end{eqnarray}
where $\nu$ is the neutrino energy in the fluid frame, $\int\intd{\Omega}$ denotes integrals over solid angle in momentum space, and
\begin{equation}
\label{eq:m1nu-momentum}
p^\alpha = \nu\left(u^\alpha + l^\alpha\right),
\end{equation}
where $l^\alpha u_\alpha = 0$ and $l^\alpha l_\alpha = 1$. We also make use of the decomposition of the neutrino radiation stress-energy tensor as observed by a normal observer,
\begin{equation}
\label{eq:m1nu-se-tensor-fluid}
{T_{\rm rad}}\ind{^\mu^\nu} = E n^\mu n^\nu + F^\mu n^\nu + F^\nu n^\mu + P\ind{^\mu^\nu},
\end{equation}
with $F^\mu n_\mu = P\ind{^\mu^\nu} n_\mu = F^t = P\ind{^t^\nu} = 0$. Additionally, for each species of neutrino we consider the number current density:
\begin{equation}
\label{eq:m1nu-def-num-density}
N^\mu = Nn^\mu + \mathcal{F}^\mu,
\end{equation}
where $N$ is the neutrino number density, and $\mathcal{F}^\mu$ is the number density flux. The decomposition of $N^{\mu}$ relative to the fluid frame can be expressed in terms of $J$, $H^\mu$, and the fluid-frame average neutrino energy $\langle\nu\rangle$ as
\begin{equation}
\label{eq:m1nu-def-num-density-fluid-frame}
N^\mu = \frac{Ju^\mu + H^\mu}{\langle\nu\rangle}.
\end{equation}

We define a projection operator onto the reference frame of an observer comoving with the fluid,
\begin{equation}
\label{eq:m1nu-inertial-proj}
h\ind{_\alpha_\beta} = g\ind{_\alpha_\beta} + u_\alpha u_\beta.
\end{equation}
This allows us to then use the fluid-frame variables to write equations for the energy, flux, and stress tensor in the normal frame (i.e. the frame with 4-velocity equal to the normal vector)
\begin{eqnarray}
\label{eq:m1nu-e}
E &=& W^2 J + 2W v_\mu H^\mu + v_\mu v_\nu S\ind{^\mu^\nu}, \\
\label{eq:m1nu-f}
F_\mu &=& W^2 v_\mu J + W\left(g\ind{_\mu_\nu} - n_\mu v_\nu\right)H^\nu \\
\nonumber
   & &\quad+ W v_\mu v_\nu H^\nu + \left(g\ind{_\mu_\nu} - n_\mu v_\nu\right) v_\rho S\ind{^\nu^\rho}, \\
\label{eq:m1nu-p}
P\ind{_\mu_\nu} &=& W^2 v_\mu v_\nu J + W\left(g\ind{_\mu_\rho} - n_\mu v_\rho\right) v_\nu H^\rho \\
\nonumber
          & &\quad+ \left(g\ind{_\mu_\rho} - n_\mu v_\rho\right)\left(g\ind{_\nu_\kappa} - n_\nu v_\kappa\right)S\ind{^\rho^\kappa} \\
\nonumber
          & &\quad+ W\left(g\ind{_\rho_\nu} - n_\rho v_\nu\right)v_\mu H^\rho,
\end{eqnarray}
by making use of the decomposition of the 4-velocity, $u^\mu = W\left(n^\mu + v^\mu\right)$.

Evolution equations for $\tilde{E} = \sdetg E$, $\tilde{F}^i = \sdetg F^i$, and $\tilde{N} = \sdetg N$ can then be written in conservative form:
\begin{eqnarray}
\label{eq:m1nu-e-evolution}
\partial_t \tilde{E} &+& \partial_j \left(\alpha \tilde{F}^j - \beta^j \tilde{E} \right) = \\
\nonumber
& &\alpha \left(\tilde{P}\ind{^i^j} K\ind{_i_j} - \tilde{F}^j \partial_j \ln \alpha - {\tilde{S}_{\rm rad}}^\alpha n_\alpha \right) \\
\label{eq:m1nu-f-evolution}
\partial_t \tilde{F}_i &+& \partial_j \left(\alpha \tilde{P}\ind{_i^j} - \beta^j \tilde{F}_i \right) = \\
\nonumber
&-& \tilde{E} \partial_i \alpha + \tilde{F}_k \partial_i \beta^k + \frac{\alpha}{2} \tilde{P}\ind{^j^k} \partial_i \gamma\ind{_j_k} + \alpha {\tilde{S}_{\rm rad}}^\alpha \gamma\ind{_i_\alpha}, \\
\label{eq:m1nu-n-evolution}
\partial_t \tilde{N} &+& \partial_j \left(\alpha \sdetg \mathcal{F}^j - \beta^j \tilde{N} \right) = \alpha \sdetg C_{(0)},
\end{eqnarray}
where $\tilde{P}\ind{_i_j} = \sdetg P\ind{_i_j}$. Complete treatment of these equations requires prescriptions for the closure relation that computes $P\ind{^i^j}(E,F_i)$, the computation of $ \mathcal{F}^j$ and the collisional source terms ${\tilde{S}_{\rm rad}}^\alpha$ and $C_{(0)}$ which couple the neutrinos to the fluid (and introduce corresponding source terms to the right hand side of Eq. \ref{eq:fluid-energy-evolution}, \ref{eq:fluid-momentum-evolution}, and \ref{eq:fluid-composition-evolution}). Details on the treatment for these are beyond the scope of this paper, and are available in \cite{Foucart2015,Foucart2016}.

One notable detail of our moment scheme that has to be handled carefully in axisymmetry, however, it the treatment of high-opacity regions. As written above, the two-moment equations lead to excessive diffusion in that regime. When the optical depth of a grid cell becomes $\geq 1$, it would be more accurate to switch to a one-moment scheme ('M0'), with a closure set by the known value of the momentum density and pressure tensor in that regime:
\begin{equation}
H_\mu^{\rm M0} = \frac{1}{3\kappa_t} \partial_\mu J^{\rm M0};\,\, S^{\rm M0}_{\mu\nu}=\frac{1}{3} J^{\rm M0} \left(g_{\mu\nu}+u_\mu u_\nu\right)
\end{equation}
with $\kappa_t$ the total opacity of the fluid to neutrinos (absorption and scattering) and
\begin{equation}
J^{\rm M0} =  \frac{3E}{4W^2-1}.
\end{equation}
We follow a slight modification of the scheme proposed in~\cite{Audit2002}, and instead correct the numerical fluxes (divergence terms) in the evolution equations so that $\tilde E,\tilde N$ evolve as solutions of the diffusion equation in the limit of high $\kappa_t$. Let us assume that $F^{\rm M1}_{\tilde E},F^{\rm M1}_{\tilde F}, F^{\rm M1}_{\tilde N}$ are the numerical fluxes in the two-moment scheme (calculated using our standard closure $P(E,F^i)$ and the HLL Riemann solver), and $F^{\rm M0}_{\tilde E},F^{\rm M0}_{\tilde F}, F^{\rm M0}_{\tilde N}$ are the same fluxes calculated using the 'M0' closure (i.e. calculating $\tilde F,\tilde P$ from $J^{\rm M0},H^{\rm M0},S^{\rm M0}$). We use as numerical fluxes
\begin{equation}
\label{eq:asymp_EN}
  F_{\tilde E,\tilde N} = aF^{\rm M1}_{\tilde E,\tilde N} + (1-a)F^{\rm M0}_{\tilde E,\tilde N} 
\end{equation}
with $a=\min{(1,\tanh{A})}$, $A=(\kappa_t \Delta x)^{-1}$, and
\begin{equation}
  \label{eq:asymp_F}
F_{\tilde F} = \tilde A^2 F^{\rm M1}_{\tilde F} + (1-\tilde A^2)F^{\rm M0}_{\tilde F},
\end{equation}
with $\tilde A=\min{(1,A)}$.

Numerical implementation of these M0 fluxes must be done with care.  Our numerical methods are designed for conservation-type equations, and the evolution of fields at cell centers is calculated from source terms and from the flux into and out of cells at the faces where they intersect adjacent cells.  Thus,
the numerical fluxes have to be estimated on cell faces (halfway between grid points).  In fact, we reconstruct flux values on each side of a face (to be combined by the approximate Riemann solver for a shock-capturing scheme). 
Terms linear in $J^{\rm M0},S^{\rm M0}$ are {\it advection} and {\it pressure gradient terms} that can be computed either using the 'left' or 'right' state of $E$ on a face. If both states agree on the sign of the advection speed, we use the upstream value of $E$ to calculate these terms. If they do not, we set all advection/pressure terms to zero. Terms linear in $H^{\rm M0}$, on the other hand, are {\it diffusion terms} that require the knowledge of $\partial_\mu J^{\rm M0}$ on cell faces. On a cell in direction '$\mu$', this can easily be estimated from the value of $J^{\rm M0}$ at neighboring cell centers. For other directions, we (a) calculate $\partial_\mu J^{\rm M0}$ on cell edges by averaging its value on the neighboring faces where it can be evaluated using simple finite differencing; and (b) calculate $\partial_\mu J^{\rm M0}$ on the cell faces where the simple finite differencing method does not work by taking the smallest value of $|\partial_\mu J^{\rm M0}|$ on neighboring cell edges (if both neighbors agree on the sign of the derivative), or setting it to zero (if they do not agree on that sign). This method is inspired from the treatment of derivatives entering the viscous stress tensor in Parrish {\it et al}~\cite{2012MNRAS.422..704P}.

\subsubsection{Viscosity}
\label{sec:viscosity}

Viscosity is implemented using the approach of \cite{Radice2017} that extends the Newtonian large-eddy simulation framework to general relativistic systems. In the large-eddy simulation framework, we recognize that although the equations for energy and momentum evolution allow for evolving modes at all scales, in numerical simulations on a discrete grid we can only evolve modes for which we have sufficient resolution to cover. Thus each computational cell deals with averaged values, while any modes smaller than the cell are removed.

We therefore average over and filter out small scales in the velocity field, leaving equations for the resolved fields:
\begin{eqnarray}
\label{eq:visc-energy-evolution}
\partial_t \overline{\tau} &+& \partial_j \left(\overline{\tau {v_T}^j} + P \sdetg \alpha \overline{v^j} \right)
 = \alpha \sdetg \left( K\ind{_i_j} \overline{S\ind{^j^k}} - \overline{S^i} \partial_i \log \alpha \right) \\
\label{eq:visc-momentum-evolution}
\partial_t \overline{S_i} &+& \partial_j \left(\overline{S_i {v_T}^j} + \alpha P \sdetg \delta\ind{_i^j} \right) = \\
& &
\nonumber
\alpha \sdetg \left( \frac{1}{2} \overline{S\ind{^j^k}} \partial_i \gamma\ind{_j_k} + \frac{1}{\alpha} \overline{S_k} \partial_i \beta^k - \frac{\left(\overline{\tau} + \overline{\rho_*}\right)}{\sdetg} \partial_i \log \alpha \right),
\end{eqnarray}
where $K\ind{^i^j}$ is the extrinsic curvature and $S\ind{^i^j} = S^i v^j + P \gamma\ind{^i^j}$.  For our simulations, we will take the averages to be cell averages.  In order to complete this set of mean-field equations, we must provide a closure condition for the quantity $\overline{S^i v^j}$:
\begin{equation}
\label{eq:visc-closure}
\overline{S_i v_j} = \overline{S_i} \overline{v_j} + \tau\ind{_i_j}.
\end{equation}
$\tau\ind{_i_j}$ is the subgrid scale stress tensor, that captures the turbulent modes unresolved by our grid. We model this tensor using
\begin{equation}
\label{eq:visc-tau}
\tau\ind{_i_j} = -2 \nu_T \rho h W^2 \left[ \frac{1}{2} \left(\nabla_i \overline{v_j} + \nabla_j \overline{v_i} \right) - \frac{1}{3} \nabla_k \overline{v^k} \gamma_{ij} \right],
\end{equation}
where $\nabla$ is the covariant derivative compatible with $\gamma\ind{_i_j}$. The quantity $\nu_T$ possesses a dimension of a viscosity, which leads us to the assumption
\begin{equation}
\label{eq:visc-mixing-length}
\nu_T = \lmix c_s
\end{equation}
where $c_s$ is the sound speed of the local fluid. $\lmix$ is the characteristic length over which our subgrid scale turbulence occurs and is known as the mixing length.

As explained in~\cite{Duez:2020}, we find that, to maintain the relations Eq.~\ref{eq:fluid-density}--~\ref{eq:fluid-momentum} for resolved fields, Eq.~\ref{eq:visc-energy-evolution} must be altered.  In this paper, we use the energy equation with the correction to 2nd order in $v$, which is
\begin{eqnarray}
\label{eq:visc-energy-evolution-added-term}
\partial_t \overline{\tau} &+& \partial_j \left(\overline{\tau} \overline{v_T{}^j} + P \sdetg \alpha \overline{v^j} \right) = \\
\nonumber
& &
\alpha \sdetg \left( K\ind{_i_j} \overline{S\ind{^j^k}} - \overline{S^i} \partial_i \log \alpha \right) - \partial_j \left(\sdetg \tau\ind{^j^k} \overline{v}_k \right).
\end{eqnarray}

\subsection{Multipatch Axisymmetry}

Multipatch methods work by dividing the computational domain into separate domain patches, each of which may have its own local coordinate system $x^i_L$ related to the global coordinate system $x^i_G$ by a map which controls the embedding of the domain in global space. In local coordinates, the patch is (for all applications in this paper) a simple Cartesian grid. The basis vectors $\partial/\partial x^i_L$ and $\partial/\partial x^i_G$ are then related by the Jacobian transformation matrix of the map. Importantly, the patches may have differing shapes in the global coordinate system that can be tailored to better capture the desired features of the simulation. Since our evolution equations for the conservative variables are generally covariant, evolution can be performed directly in the local coordinate system of each individual patch and then the result can be transformed back to the global coordinate system for any necessary communication of information between patches.

SpEC is parallelized using MPI, and the division of grids into patches is used to divide the simulation work.  Each domain patch is assigned to a particular processor.  For the simulations in this paper, we choose to assign one fluid patch to each processor and, for runs with concurrent spectral evolution of the metric, one pseudospectral patch.

Communication between domain patches occurs through synchronizing values in the ghost zones of each patch at the end of each timestep. In the case that these subdomain patches overlap but do not have directly matching points we communicate data by interpolating values between points.  When evolving a vector potential $A_i$, an additional ghost zone synchronization must be carried out on the magnetic field after it is computed from the curl of $A_i$.  In this case, only the outermost layer of ghost zone points, not a full stencil, should be synchronized.  Our curl operator is second-order and only uses nearest neighbors, so interpolated synchronization of additional layers results in magnetic monopole artifacts that grow quickly~\cite{Nouri2018}.

Additionally, we create ghost zone points that extend beyond any symmetry boundaries that we have defined in order to impose boundary conditions. During the communication phase, these ghost zone points are filled with data from the live points using the appropriate symmetry conditions.  In particular, we must impose symmetry conditions on the symmetry axis.  Let us define basis vectors as follows.  Imagine a 2D plane, which will represent the computational domain, that crosses the symmetry axis and introduce Cartesian coordinates and basis vectors on the plane.  The direction parallel to the axis is $\partial_z$.  The cylindrical radius giving the coordinate distance to the axis is called $\varpi$, and the corresponding Cartesian coordinate on the plane is $\varpi_C$.  Set $\varpi_C=\varpi$ when $\varpi_C>0$, but on the other side of the axis, $\varpi_C=-\varpi$.  Finally, there is a third Cartesian axis $\partial_y$ points out of the plane and {\it on the plane} is related to the azimuthal direction $\partial_y=\varpi_C^{-1}\partial_{\phi}$.  To impose the axisymmetry condition, we add a stencil of ghost zones across the axis at negative $\varpi$.  Note that $\partial_{\varpi_C}$ and $\partial_{\varpi}$ are antiparallel in the ghost zone region, as are $\partial_y$ and $\partial_{\phi}$.  We use cell-centered grids in the $\varpi_C$ direction, so the first live point has center half a grid spacing offset from the axis, and no point (live or ghost) is centered exactly on the axis.  For scalar quantities, the axisymmetry condition is $f(-\varpi)=f(\varpi)$.  For vectors, $v^{\varpi_C}(-\varpi_C)=v^{\varpi_C}(\varpi_C)$, $v^z(-\varpi_C)=v^z(\varpi_C)$, $v^{y}(-\varpi_C)=-v^{y}(\varpi_C)$.  Below, we will ignore the distinction between $\varpi$ and $\varpi_C$, since it is only relevant for ghost zones.

\begin{figure}
\centering
\includegraphics*[width=.7\textwidth]{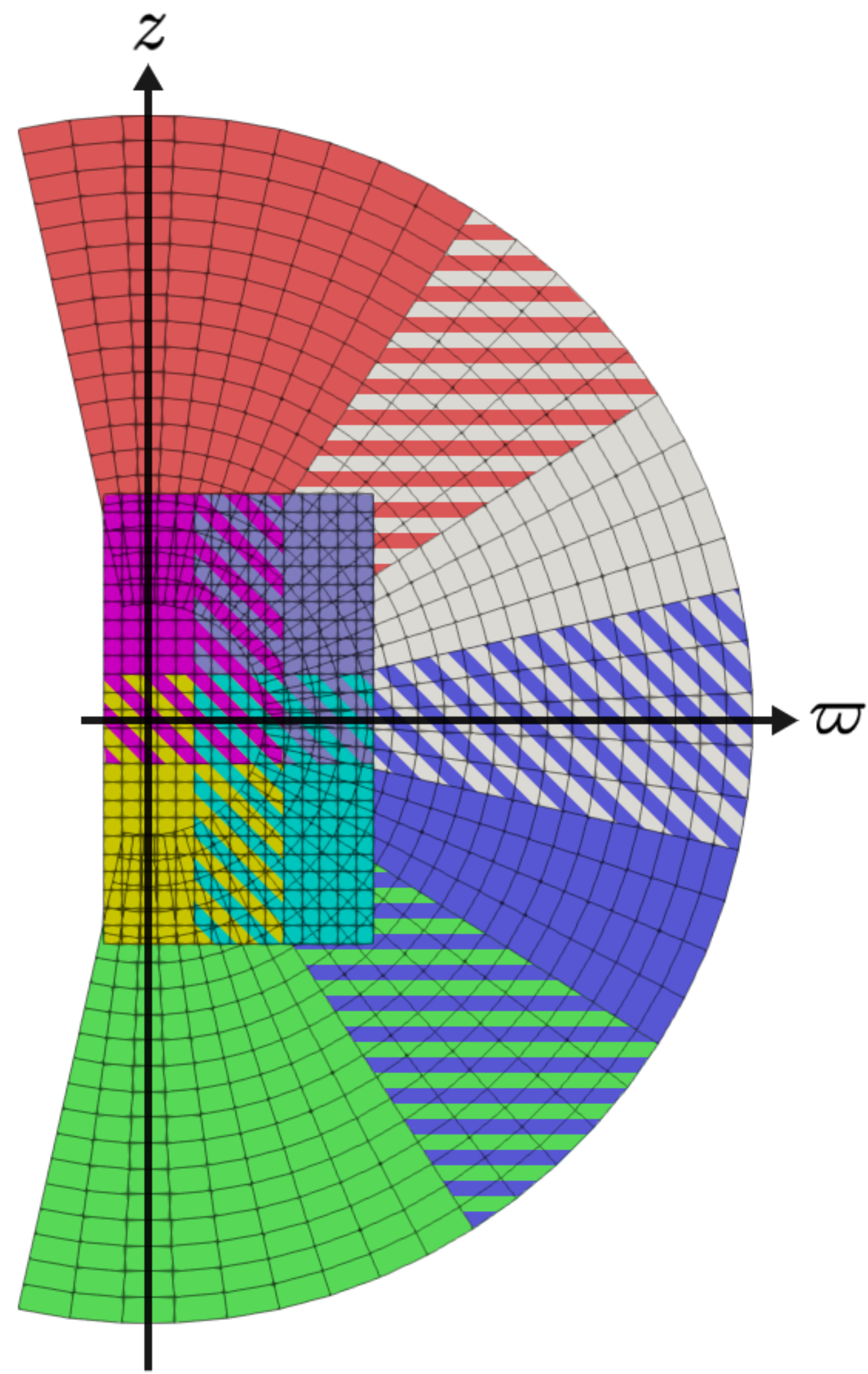}
\caption{Example of a 2D multipatch grid, for use with axisymmetry, composed of overlapping square (cylindrical-polar) and wedge (spherical-polar) grid shapes. Grid points are arranged so that the coordinate singularity at the symmetry axis falls between grid points. Points extending beyond the symmetry axis are ghost zone points used to impose boundary conditions. Striped regions show portions of the grid where two or more patches are overlapping with matching points.}
\label{fig:multipatch-example}
\end{figure}

When evolving a three-dimensional system using a two-dimensional computational domain, each gridpoint represents a ring labeled by two nonazimuthal coordinates. Quite general 2D maps are possible to relate local to global coordinates, but two are particularly useful. A linear map ($x^i_G=a_i x^i_L + b_i$) corresponds to patches that are globally rectangular blocks, covering cylinders in 3D. A polar map [e.g. $x^1_G=x^1_L\cos(x^2_L)$, $x^2_G=x^1_L\sin(x^2_L)$] corresponds to patches that are globally wedges of circles, covering a specified range of polar $r$, $\theta$. A combination of wedges covering $0<\theta<\pi$ in 2D covers a spherical shell domain in 3D. A general 2D grid can contain arbitrary combinations of rectangular blocks and wedges, as shown in Fig. \ref{fig:multipatch-example}.

Although the grid is 2D, the tangent space on which vectors live is still 3D; even axisymmetric systems can have azimuthal velocity and magnetic field components, for example. The third coordinate in the local coordinate system is set to be the global azimuthal $\phi$. Then the local coordinates for a rectangular block will be (up to linear transformation) cylindrical-polar, while the local coordinates for a wedge patch will be (up to linear transformation) spherical-polar. By modifying the map Jacobian, we can make the existing transformation between local and global coordinates handle transforming the third coordinate into an azimuthal coordinate that can be used to perform axisymmetric evolutions. To do this we expand the elements of the Jacobian matrix using the chain rule to add in the effects of the polar transformation:
\begin{equation}
\label{eq:jacobian-def}
\tensor{J}{^i_j} = \pdv{x_{G}^i}{x_{L}^j} = \pdv{x_{G}^i}{x_{A}^n} \pdv{x_{A}^n}{x_{L}^j},
\end{equation}
where $x_G$ are the global coordinates, $x_L$ are the local coordinates of a given grid patch, and $x_A$ are a set of global polar coordinates. Since the global and polar coordinates only differ in terms involving the azimuthal direction, the final change from the original Jacobian, $J\indices{^i_j}$, to the new axisymmetry Jacobian, ${J_{\rm axi}}\indices{^i_j}$, will be straightforward:

\begin{equation}
\label{eq:jacobian-change}
J\indices{^i_j} =
\left(
\begin{array}{ccc}
\pdv{x_{G}^1}{x_{L}^1} & \pdv{x_{G}^1}{x_{L}^2} & 0 \\ 
\pdv{x_{G}^2}{x_{L}^1} & \pdv{x_{G}^2}{x_{L}^2} & 0 \\ 
0 & 0 & 1\\
\end{array}\right)
\rightarrow
{J_{\rm axi}}\indices{^i_j} =
\left(\begin{array}{ccc}
\pdv{x_{G}^1}{x_{L}^1} & \pdv{x_{G}^1}{x_{L}^2} & 0 \\ 
\pdv{x_{G}^2}{x_{L}^1} & \pdv{x_{G}^2}{x_{L}^2} & 0 \\ 
0 & 0 & \varpi \\
\end{array}\right),
\end{equation}
where $\varpi$ is the coordinate distance from the rotational symmetry axis and we have chosen coordinate directions 1 and 2 to correspond to the two coordinates defined by our two-dimensional computational domain and coordinate direction 3 is transformed to the axisymmetric azimuthal direction $\phi$. We also make use of the Hessian matrix in the transformation of the derivatives of metric-related quantities to the local coordinates, and must likewise make similar adjustments to the Hessian:
\begin{equation}
\label{eq:hessian-def}
\tensor{H}{^i_j_k} = \pdv{}{x_{L}^j}\left(\pdv{x_{G}^i}{x_{L}^k}\right) = \pdv{}{x_{L}^j}\left(\pdv{x_{G}^i}{x_{A}^n} \pdv{x_{A}^n}{x_{L}^k}\right).
\end{equation}
Explicitly,
\begin{eqnarray}
  \tensor{H}{^3_3_1} &=& \tensor{H}{^3_1_3} = \tensor{J}{^2_1}, \\
  \tensor{H}{^3_2_3} &=&  \tensor{J}{^2_2}, \\
  \tensor{H}{^2_3_3} &=&  -\varpi.
\end{eqnarray}

Generally the evolution of Einstein's equations using SpEC's pseudospectral grid tends to use much less computing time than the hydrodynamics evolution, so our axisymmetry method is primarily aimed at implementing axisymmetric evolution on the hydrodynamics grid while evolving Einstein's equations in 3D.  For spherical shell pseudospectral domains, whose colocation points correspond to an expansion of functions in terms of spherical harmonics, azimuthal information can be reduced by reducing azimuthal resolution, corresponding to a lowering of the azimuthal mode number $m$ retained in spectral expansions. It cannot be lowered to $m_{\rm max}=0$ because the spectral evolution uses Cartesian components of tensors. We find, however, that the speed increase from doing so is modest, and the resulting spectral grids are more prone to constraint-violating instabilities, so we have not used azimuthal resolution reduction on pseudospectral grids for the simulations in this paper.

Information required by the pseudospectral grid from the hydrodynamics grid is expanded back to 3D during communication.  2D planar fluid data can be extended to 3D for the metric source terms by rotating appropriately about symmetry plane.  For the metric data needed for the fluid evolution, as long as the pseudospectrally evolved metric nearly respects the axisymmetry, it is sufficient to take metric data from the $xz$ plane.  Deviations from axisymmetry in the evolved spacetime functions could arise from three sources.  The first is numerical error.  We have found this small enough to be ignorable, but it could artificially be removed when using spherical shell domains by filtering modes in $m$.  Alternatively, azimuthally-averaged values could be used for the fluid evolution, but we would find the idea of significant uncontrolled violation of the axisymmetry assumption troubling.  Second, even if the spacetime retains an azimuthal Killing vector, the coordinate system might evolve in a way that breaks the assumed coordinate form of axisymmetry.  Hopefully, the gauge choice will minimize such effects.  Merger simulations in SpEC use the damped harmonic gauge~\cite{szilagyi2014key}, and we have found that metric profiles following black hole-neutron star mergers are indeed nearly axisymmetric in the evolved coordinates.  This gauge also maintains axisymmetry for the live-metric single-star problem described in Section~\ref{sec:diff-star} below.  Finally, the metric could conceivably be subject to some violent non-axisymmetric instability, in which case the methods of this paper are obviously inappropriate. 

The conservative form of radiation magnetohydrodynamics evolves variables that are densities and thus proportional to $\sqrt{\gamma}$. Under local to global transformation, the metric determinant transforms as $\sqrt{\gamma_L} = J\sqrt{\gamma_G}$, where $J$ is the determinant of the Jacobian. Note that $J$ is zero on the axis, and indeed would naturally change sign there because the orientation of the basis vectors switches there. SpEC always takes a positive square root, but the only points on the other side of the axis are ghost zone points (needed to impose the symmetry boundary conditions), and non-smooth functions like $\sqrt{\gamma}$ are not interpolated or reconstructed.

\begin{figure*}
\subfloat[Without flux factoring]{
 \includegraphics[width=0.36\textwidth]{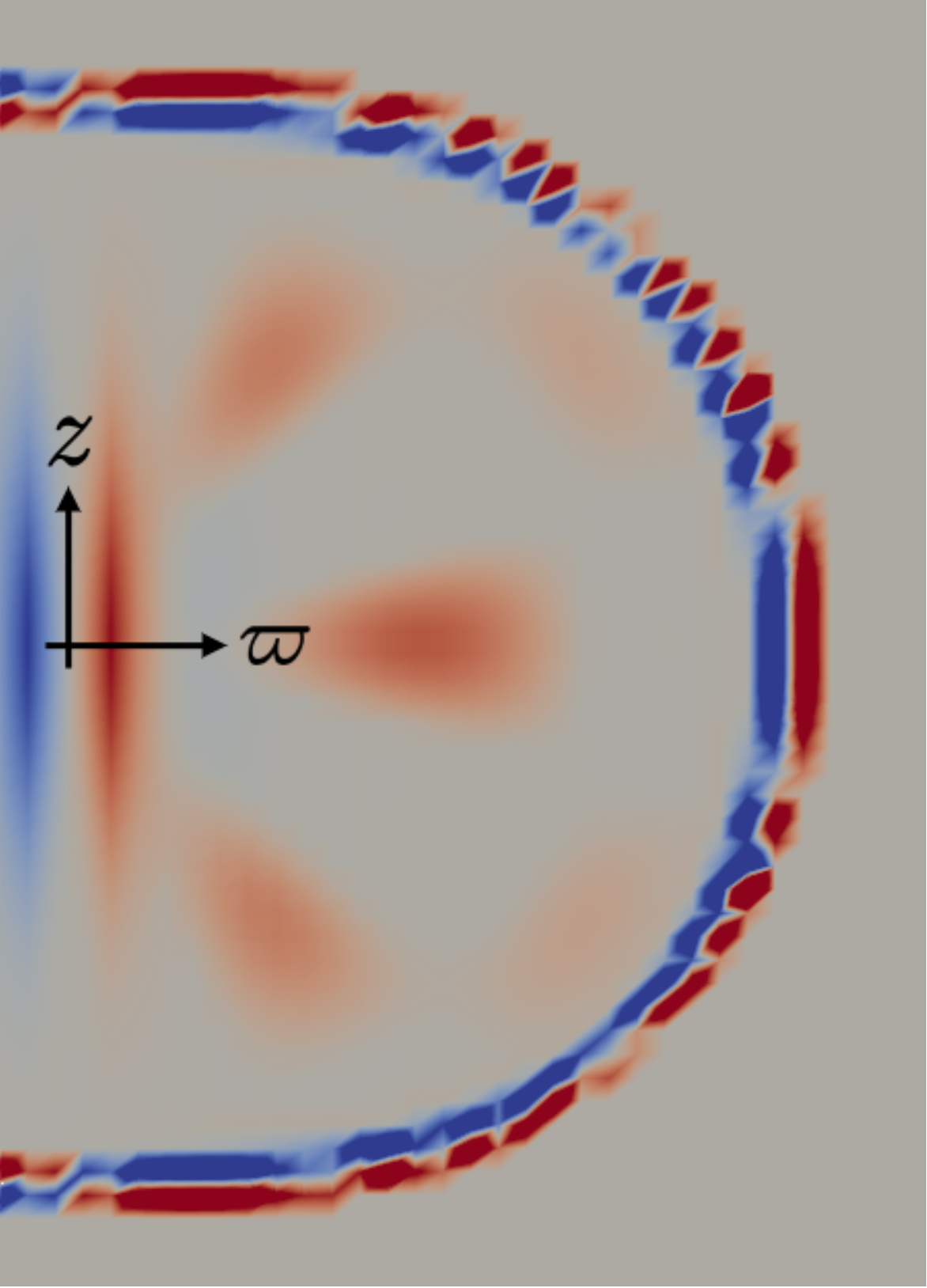}
}%
\hspace{0.1\columnwidth}%
\subfloat[With flux factoring]{
 \includegraphics[width=0.36\textwidth]{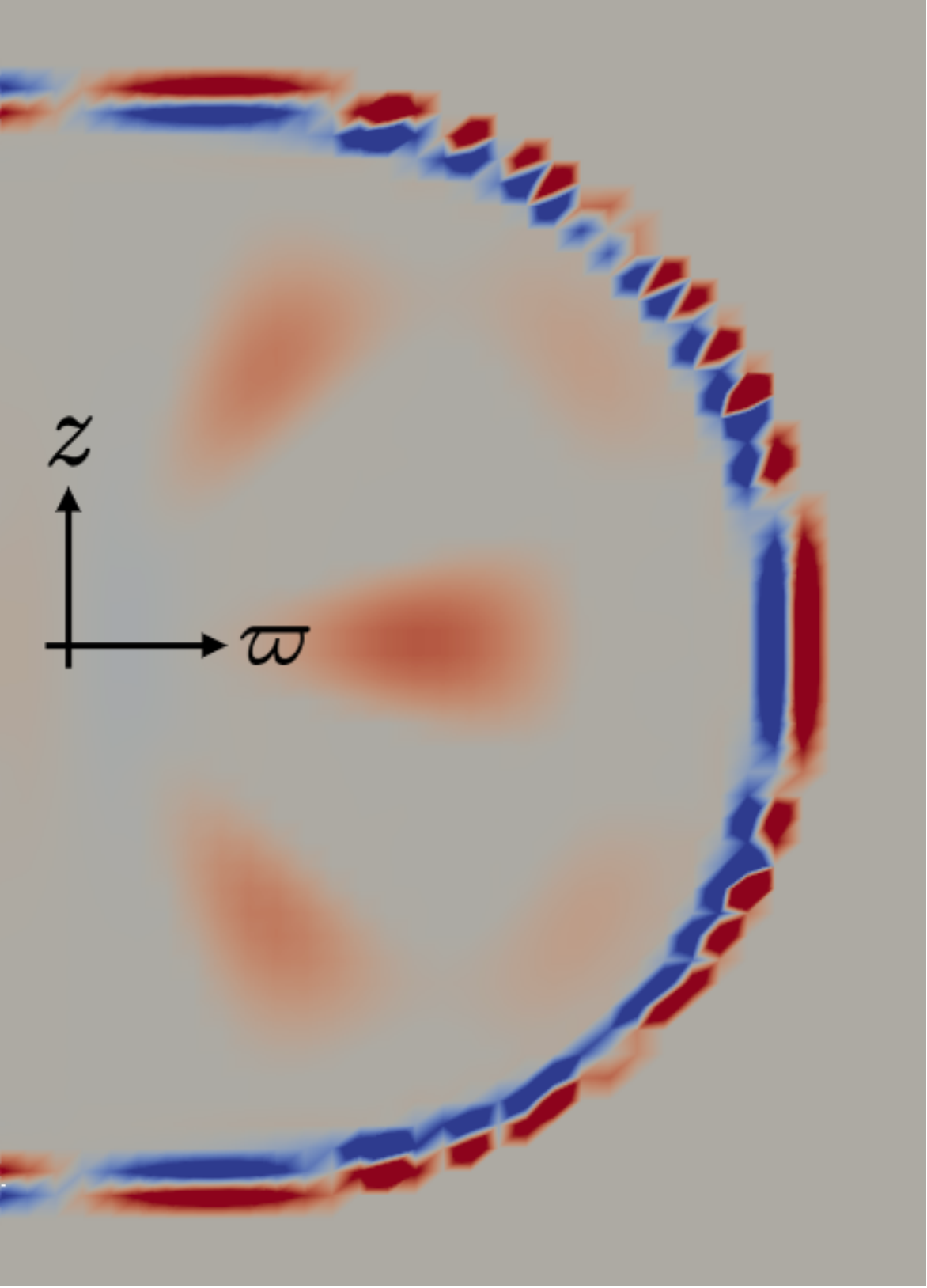}
}%
\hspace{0.025\columnwidth}%
\subfloat{
 \includegraphics[width=0.1\textwidth]{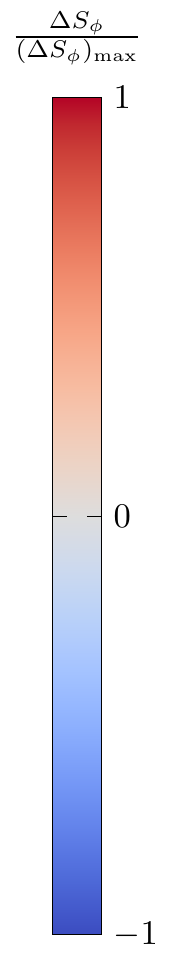}
}%
\caption{Example of error growth in the evolution of $S_\phi$ (after tranformation back to the global coordinate system) near the symmetry axis of a low resolution, differentially rotating star in a stationary state. Both images plot the difference of $S_\phi$ between the initial state and the end of the first time step, with the same color scale used for both images chosen to enhance the appearance of errors inside the star. Since the initial conditions are an equilibrium state, all deviations from zero are due to numerical error. A good handling of the symmetry axis leads to errors not being particularly large there. The left image shows the multipatch axisymmetry method applied without factoring of flux terms, while the right shows the star with factoring enabled.}
\label{fig:axisym-axis-error}
\end{figure*}

Unfortunately, when evolving, this method is prone to producing errors near the symmetry axis that, without correction, grow over time. Vector and tensor valued quantities are most heavily affected due to direct transformation of components in the azimuthal coordinate direction introducing singular terms. An example of this type of error is shown in Fig. \ref{fig:axisym-axis-error}. Eventually though, all of our evolved quantities, including scalar quantities, will suffer from errors due to also picking up a singular term in the determinant of the 3-metric.

The problem primarily occurs during the computation of the divergence of the flux term, $F_A$, in the evolution equation of a given quantity $A$
\begin{equation}
\label{eq:generic-evolution}
\partial_t A + \partial_i F_A{}^i = S_A
\end{equation}
with $S_A$ being any source terms appearing on the right-hand side of the equation.

Some early 2D general relativistic hydrodynamic simulations stabilized the axis evolution using dissipation~\cite{shibata:2000,duez:04}. 
Our solution, inspired by~\cite{Baumgarte2015}, is to factor out singular terms that have been introduced to $F_A$ during the transformation to the local coordinates prior to computing the divergence. Depending on the specific component of the flux $F_A$ corresponding to $A$, there may be multiple factors of $\varpi$ that need to be removed:
\begin{equation}
\label{eq:flux-def}
F_A{}^i = \varpi^n \tilde{F}_A{}^i,
\end{equation}
where $\tilde{F}_A$ is just the $\varpi$-factored form of the flux, and the integer $n$ will depend on $A$.  We can now instead take the divergence of this factored form of the flux and apply the chain rule, which gives
\begin{equation}
\label{eq:divergence-factoring}
\partial_i (\varpi^n \tilde{F}_A{}^i)
 = \varpi^n \partial_i\tilde{F}_A{}^i + n \varpi^{n-1} \pdv{\varpi}{x_L^i} \tilde{F}_A{}^i.
\end{equation}

We can also take advantage of the property that if the coordinate specified by $\varpi$ corresponds to one of the directions in the global coordinate system, for example if the global coordinates are Cartesian, the derivatives of $\varpi$ with respect to the local coordinates can be directly taken from components of the Jacobian dealing with the direction associated with $\varpi$. With this, all of the singular terms introduced from the polar Jacobian are removed from the divergence. Importantly though, the divergence of $\tilde{F}_A$ in the first term on the right side of this equation will need to be computed using the value of $\tilde{F}_A$ at cell faces using the Riemann solver, while $\tilde{F}_A$ in the second term on the right side will use the value at cell centers.

Additionally, since all components have now been transformed into a polar coordinate system, from the definition of axisymmetry we have
\begin{equation}
\label{eq:axisym-def}
\partial_\phi F_A{}^\phi = 0,
\end{equation}
where the $\phi$-index indicates the coordinate of the axisymmetric azimuthal direction. This allows us to ignore the azimuthal portion of the divergences so that we only need to apply the factoring to the two components of the flux that lie in the plane of the computational grid ($i$ = 1 and 2 in the below factoring).

All of our evolved quantities carry a factor of $\sdetg$, which will also acquire a singular term, from the transformation of $\gamma\ind{_i_j}$ to the local coordinate system, that also needs to be handled analytically. The flux factoring thus falls into three broad categories for our current evolution equations. Factoring of fluxes for scalar density quantities [Eq. \ref{eq:fluid-density-evolution}, \ref{eq:fluid-energy-evolution}, \ref{eq:fluid-composition-evolution}, \ref{eq:mag-scalar-potential-evolution}, \ref{eq:mag-div-clean-psi-evolution}, \ref{eq:m1nu-e-evolution}, \ref{eq:m1nu-n-evolution}, and the added term in \ref{eq:visc-energy-evolution-added-term}], takes the form
\begin{eqnarray}
\label{eq:scalar-factoring}
{F_A}^i &=& \varpi \tilde{F_A}^i, \\
\label{eq:scalar-divergence}
\partial_i {F_A}^i &=& \varpi \partial_i \tilde{F_A}^i + \pdv{\varpi}{x_L^i} \tilde{F_A}^i.
\end{eqnarray}
Factoring for covariant vector density quantities [Eq. \ref{eq:fluid-momentum-evolution}, \ref{eq:mag-vector-potential-evolution}, and \ref{eq:m1nu-f-evolution}], takes the form
\begin{eqnarray}
\label{eq:covariant-vec-factoring}
{F_A}\ind{_j^i} &=&
	\left\{\begin{array}{ll}
		\varpi \tilde{F_A}\ind{_j^i}& \text{, for } j \neq \phi \\
		\varpi^2 \tilde{F_A}\ind{_j^i}& \text{, for } j = \phi,
	\end{array}\right. \\[1em]
\label{eq:covariant-vec-divergence}
\partial_i {F_A}\ind{_j^i} &=&
	\left\{\begin{array}{ll}
		\varpi \partial_i \tilde{F_A}\ind{_j^i} + \pdv{\varpi}{x_L^i} \tilde{F_A}\ind{_j^i}& \text{, for } j \neq \phi \\[0.8em]
		\varpi^2 \partial_i \tilde{F_A}\ind{_j^i} + 2 \varpi \pdv{\varpi}{x_L^i} \tilde{F_A}\ind{_j^i}& \text{, for } j = \phi.
	\end{array}\right.
\end{eqnarray}
Factoring for contravariant vector density quantities [Eq. \ref{eq:mag-div-clean-b-evolution}], takes the form
\begin{eqnarray}
\label{eq:contravariant-vec-factoring}
{F_A}^{j i} &=&
	\left\{\begin{array}{ll}
		\varpi \tilde{F_A}\ind{^j^i}& \text{, for } j \neq \phi, \\
		\tilde{F_A}\ind{^j^i}& \text{, for } j = \phi,
	\end{array}\right. \\[1em]
\label{eq:contravariant-vec-divergence}
\partial_i {F_A}\ind{^j^i} &=&
	\left\{\begin{array}{ll}
		\varpi \partial_i \tilde{F_A}\ind{^j^i} + \pdv{\varpi}{x_L^i} \tilde{F_A}\ind{^j^i}& \text{, for } j \neq \phi, \\[0.8em]
		\partial_i \tilde{F_A}\ind{^j^i}& \text{, for } j = \phi.
	\end{array}\right.
\end{eqnarray}
In each of these, the index $i$ only covers coordinates 1 and 2 due to Eq. \ref{eq:axisym-def}.  SpEC and most other relativistic hydrodynamics codes use conservative shock capturing techniques with approximate Riemann solvers.  For codes of this type, a convenient way to implement this factoring program is to use a different coordinate basis, with $\frac{1}{\varpi}\frac{\partial}{\partial x^\phi}$ instead of $\frac{\partial}{\partial x^\phi}$, on cell faces than on cell centers.  That is, one simply reconstructs factored quantities.

When evolving a magnetic vector potential, it is also necessary to factor $A_{\phi}$ when
computing $B^i$.
\begin{equation}
 \partial_{i}A_{\phi} = \varpi\partial_{i}\tilde{A}_{\phi} + \tilde{A}_{\phi}\pdv{\varpi}{x^i},
\end{equation}
where $\tilde{A_\phi} = A_\phi / \varpi$ \footnote{In fact, only factoring for the coordinate $i$ nearly parallel to $\varpi$ on the axis is necessary.}.

In their factored form, the principle part of the fluid equations matches 2D Cartesian hydrodynamics (with both sides of the equations divided by a $\varpi$ factor) and so should have similar stability properties.  A similar factoring scheme has been extensively tested in 3D~\cite{Baumgarte2013,Montero2014,Baumgarte2015}.

The neutrino variables in the M1 scheme are handled in the same way.  Even in axisymmetry, the first moment $\tilde{F}^i$ is three-dimensional, and for any transport scheme axisymmetry imposes no constraint on the momentum-space dependence of the distribution function at any spatial point.  Factoring of scalar and vector densities is carried out as above.  This requires the weighted averages of M1 and M0 fluxes from Eq.~\ref{eq:asymp_EN} and~\ref{eq:asymp_F} computed at cell centers.  The value of $\partial_{\mu}J^{\rm M0}$ on cell centers is estimated by a centered second-order finite difference using center values of neighboring cells.  Since the sign of each component of the advection speed at a cell center is always unambiguous (as opposed to cell faces, for each of which there are two reconstructions), we always add the advective and pressure gradient components of M0 fluxes at cell centers.  In fact, this contribution is needed to avoid axis artifacts.  The $\tilde{P}\ind{^j^k} \partial_i \gamma\ind{_j_k}$ source term in Eq.~\ref{eq:m1nu-f-evolution} contains a singular term (from the $\varpi^2$ factor in $\gamma\ind{_3_3}$) which is canceled by a matching term in the flux from $\alpha \tilde{P}\ind{_i^j}$ evaluated at cell centers.  In the optically thick limit, this matching term is formally in the advective and pressure gradient part of the M0 flux of $\tilde{F}^i$.  If over an extended optically thick region of the grid the advection speeds on the left and right of cell faces either vanish or differ in sign, then there can be an inconsistency between how the flux is computed at cell faces (for which the advective term would be absent) and cell centers (for which it would be present), which we find also creates axis artifacts when using non-rectangular grids.  Such a situation is not likely to occur in realistic simulations, but it does occur in the test problem in Section~\ref{sec:nutest} below, in which velocities are set to zero.  It can be dealt with in a number of ways.  One simple way is to add advective fluxes and radiation pressure gradient terms, or at least the latter, on faces even when advective speeds are zero, in that case using the average of values calculated from the two reconstructions.  Another simple way is to fall back to M1 fluxes for $\tilde{F}^i$ and handle these as in~\cite{Foucart2015}.

Metric-related quantities ($\gamma\ind{_i_j}$, $\alpha$, $\beta$) are evolved on their own separate spectral grid in 3D and are communicated to the hydrodynamics grid at the end of each time step. Spatial derivatives of these metric quantities are computed while on the metric grid and then communicated to the hydrodynamics grid, at which point they can be transformed into the local coordinate system as needed. The transformation to local coordinates uses the analytic Jacobian and Hessian, so metric derivatives automatically have their singular factors treated analytically. The transformation equations for global to local components of metric derivatives are 
\begin{eqnarray}
\label{eq:dshift}
      {\beta_L}\ind{^j_{,i}} &=& (J^{-1})\ind{^j_{\overline{j}}} J\ind{^{\overline{i}}_i} {\beta_G}\ind{^{\overline{j}}_{,\overline{i}}}
      - {\beta_L}^k(J^{-1})\ind{^j_{\overline{j}}} H\ind{^{\overline{j}}_i_k},\\
{\gamma_L}\ind{^i^j_{,k}} &=& (J^{-1})\ind{^i_{\overline{i}}} (J^{-1})\ind{^j_{\overline{j}}} J\ind{^{\overline{k}}_k} {\gamma_G}\ind{^{\overline{i}}^{\overline{j}}_{,\overline{k}}} \\
  & &
\nonumber
\quad- H\ind{^{\overline{m}}_k_n}[{\gamma_L}\ind{^n^j}(J^{-1})\ind{^i_{\overline{m}}} + {\gamma_L}\ind{^i^n}(J^{-1})\ind{^j_{\overline{m}}}].
\label{eq:dmetric}
\end{eqnarray}

\subsection{Auxiliary Entropy Variable}

After each substep, the evolved variables ($\rho_*$,$\tau$,$S_i$,$\rho_*Y_e$,$\tilde{B}^i$) must be used
to recover the primitive variables ($\rho_0$,$T$,$Y_e$,$u_i$,$B^i$), a process that involves
multi-dimensional root-finding. In particular, if the internal energy is small compared to kinetic
or magnetic energy, the temperature recovered from total energy and momentum densities will be
unreliable. Due to numerical error, recovered $T$ and $u_i$, especially at very low densities,
may be unphysical, or there may not even be a set of primitive variables corresponding to the
evolved variables at a point.

As in~\cite{Noble:2008tm,Nouri2018}, we introduce an auxiliary entropy density evolution variable $\rho_*S$, where
$S$ is the specific entropy. The variable $\rho_*S$ obeys
a continuity equation (viscous and neutrino source terms being unimportant for its purpose)
which can be treated in axisymmetry like the other scalar density evolution equations. After
each substep in time, SpEC first attempts to recover primitive variables using the standard
evolution variables. If this is not possible, or if the recovered specific entropy decreases
by more than a fixed percentage compared to its advected value~\footnote{The rationale is that
 shocks, magnetic reconnection, and viscosity can only increase entropy. If the loss of a
 significant percentage of the entropy at a gridpoint in one timestep by neutrino cooling is
 considered plausible in a given simulation, this condition would have to be relaxed, for
 instance to require only that the entropy remain positive.  For this paper, we allow the entropy to decrease to 0.97 times its value advected from the previous timestep for problems without neutrino radiation.  The 3\% buffer lets us avoid only allowing heating errors, which could lead to a systematic drift of temperature.  For the neutrino transport problem (Section~\ref{sec:nutest}), the entropy variable was not used.}, primitive variables are recovered
disregarding $\tau$ and using $\rho_*S$. At the end of each substep, the primitive variables
are used to reset all evolution variables, so that $\tau$ and $\rho_*S$ are synchronized to each
other.  The entropy variable allows more reasonable recovery of fluid internal energy at points
where this component of the energy is a subdominant contribution to $\tau$ and $S_i$.  A
particularly challenging problem is accurate primitive variable recovery in magnetically
dominated regions like jets.  (See~\cite{Noble2006,Siegel:2017sav,Kastaun:2020uxr} on inversion
methods for this case.)

For physical equations of state (e.g. finite-temperature nuclear-theory based EoS),
the actual statistical mechanical entropy per baryon can be
used to define $S$. However, in numerical relativity, equations of state are commonly used
which have no uniquely defined entropy or temperature, although with absolute zero specified
from outside (e.g. for Gamma-law EoS, a value of the polytropic constant is defined to be ``cold''). A common
case is an EoS with nuclear physics-motivated cold component plus a simple thermal Gamma-law
component added on. In terms of baryonic number density $n=\rho_0/m_{\rm amu}$ and internal
energy density $u$,
\begin{equation}
 P(n,u) = P_c(n) + (\Gamma_{\rm th}-1)(u - u_c),
\end{equation}
where
\begin{equation}
 P_c(n) = n^2\frac{d[U_c/n]}{dn}.
\end{equation}
The first law gives
\begin{equation}
 nTdS = -(u+P)dn + ndu.
\end{equation}
Combining the three above equations yields, after a short calculation,
\begin{equation}
 nTdS = \rho_0^{\Gamma_{\rm th}}d\left[(u-u_c)\rho_0{}^{-\Gamma_{\rm th}}\right],
\end{equation}
so $(u-u_c)\rho_0{}^{-\Gamma_{\rm th}}$ advects for adiabatic
change, indicating that this is an acceptable $S$ variable. For Gamma-law
EoS, one can set $u_c=0$, yielding the standard auxiliary entropy
variable (up to a scaling factor) for this case.

\subsection{Low-density treatment}

We impose a density floor in low-density regions outside stars, which is necessary to avoid division by zero in our finite difference solver.  Also, for densities $\sim$ 2 decades above the floor and below, we impose limits on temperature and velocity; see~\cite{Foucart:2013a} for details.  In the presence of a magnetic field, we do not limit the components of the velocity normal to the field lines even at the lowest densities, to avoid altering the electric field~\cite{Muhlberger2014}.  Test problems with low-density regions also use the auxiliary entropy variable to assist recovery of primitive variables from conservative variables.

\section{Tests} \label{sec:tests}

\subsection{Mass Conservation of an advected pulse}

\begin{figure}
\centering
\includegraphics[width=0.75\textwidth]{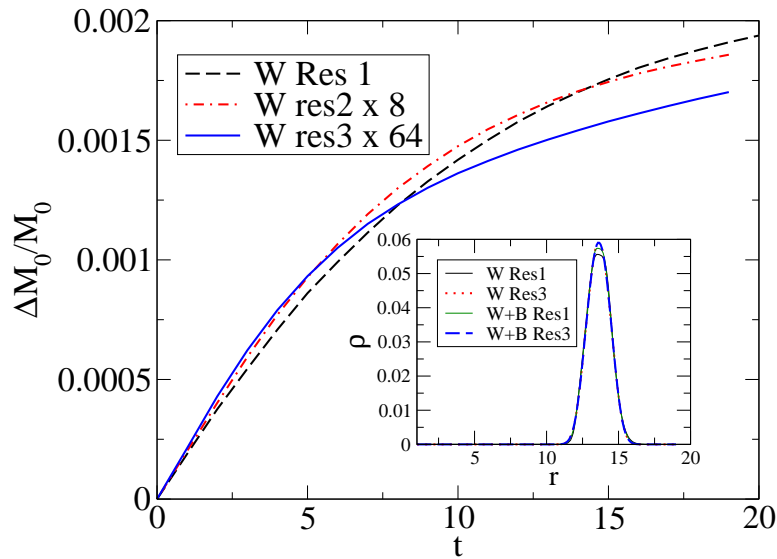}
\caption{Fractional error in rest mass $M_0$ as a function of time for 3
  resolutions of wedge grids for the outgoing density pulse problem.
  Inset:  the equatorial profile of the pulse at $t=19$ for the lowest
  and highest resolution wedge (W) and wedge plus block (W$+$B) grids.
  Our integration routine does not handle overlapping grids accurately,
  so we do not plot the time evolution of $M_0$ for W$+$B grids, but
  the final mass error (after the pulse leaves the overlap region) is
  reported in Table~\ref{tab:mass_consv}.}
\label{fig:mass-convergence}
\end{figure}

Factoring the continuity equation does lead to truncation error in rest mass
conservation.  So too does the interpolation needed to fill ghost zones at
non-matching patch boundaries for grids like that shown in
Figure~\ref{fig:multipatch-example}.  For our applications, we have found
these errors to be quite small, but we quantify them here for a simple test.
We introduce a density pulse $e^{(r-5)^2}(\cos^2\theta+1)$ moving radially
outward at 0.5$c$ in Minkowski spacetime.  We evolve until the center of the
pulse reaches $r=14$.
For a first set of runs (W), we use spherical-polar wedges covering $1<r<20$.
The lowest resolution has 100 radial points and 40 polar points; each
subsequent resolution has double the points in each direction as the previous
one.  For a second set of runs (W$+$B), we use rectangular blocks surrounded by
wedges, with the interface in the $7<r<10$ region.  The lowest resolution
has 100 radial points and 40 polar points in the wedges and $100\times 50$
points in the blocks; each subsequent resolution doubles the points in each
direction in each patch.

\begin{table}
  \begin{center}
    \begin{tabular}{c|cccccc}
      \toprule
      Grid & W Res1 & W Res2 & W Res 3 & W$+$B Res1 & W$+$B Res2 & W$+$B Res3\\
      \midrule
      $10^4\Delta M_0/M_0$ & 19 & 2.3 & 0.27 & 26 & 1.6 & 0.085 \\
      \bottomrule
    \end{tabular}
    \caption{Change in rest mass during evolution of density pulse propagation
      for pure wedge (W) and wedge plus block (W$+$B) grids.}
    \label{tab:mass_consv}
  \end{center}
\end{table}

The patch is able to probagate through the interpolation region without
picking up noticable artifacts or deviation from the purely wedge grid.  This
is not surprising, since in an earlier paper we were able to propagate a
shock through a similar grid in 3D~\cite{Nouri2018}.  In
Table~\ref{tab:mass_consv}
we report the change in rest mass between initial and final times.  We see
roughly 3rd-order convergence thanks to our 5th-order WENO
reconstruction scheme, 3rd-order ghost zome interpolation, and 3rd-order time
differencing.  This is
also demonstrated in Fig.~\ref{fig:mass-convergence}.

\subsection{TOV Star}

\begin{figure}
\centering
\includegraphics[width=0.75\textwidth]{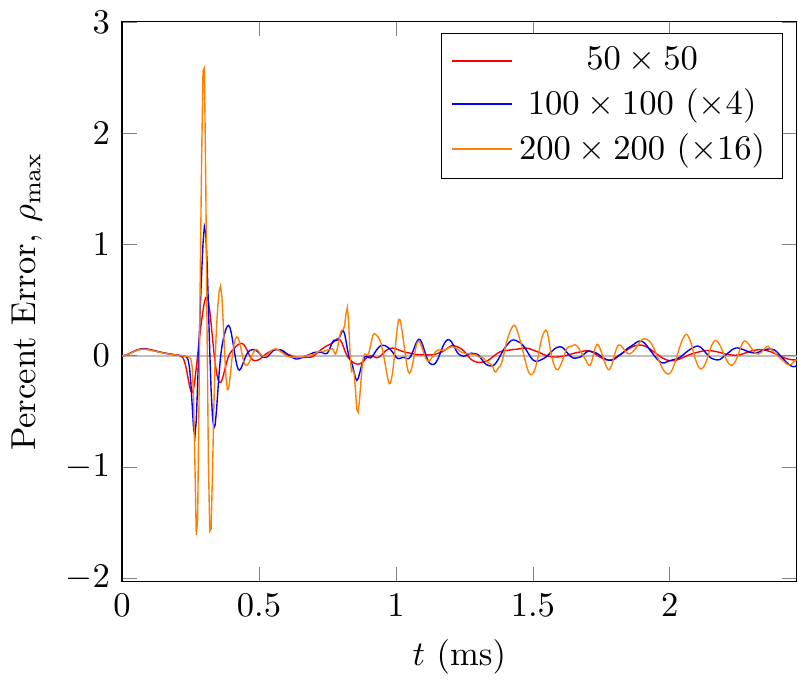}
\caption{Percent error in the maximum density for the TOV star. Error is shown for grid resolutions of $50\times50$, $100\times100$, and $200\times200$. The error for the $100\times100$ and $200\times200$ resolutions have been scaled up by the square of the change in resolution from the $50\times50$ case.}
\label{fig:tov-star-convergence}
\end{figure}

Initial stability testing was performed using a Tolman–Oppenheimer–Volkoff (TOV) star in a stationary state. The star was created using a polytropic equation of state with polytropic index $\Gamma$ = 2, polytropic constant $\kappa$ = $100G^3c^{-4}M_{\odot}^2 = 1.82\times10^{10}$~\si{cm^5.g^{-1}.s^{-2}}, and a central density of $7.72\times10^{14}$ \si{g.cm^{-3}}. This resulted in a gravitational mass of 1.38 $\Msun$, a baryonic rest mass of 1.49 $\Msun$ and a circumferencial radius of 14.22 \si{km}. The star was evolved for 2.46 \si{ms} in 2D, using both axisymmetry and equatorial symmetry.  This is 20 dynamical timescales, using for the dynamical timescale the free-fall time $\sqrt{R^3/(GM)}$.  The computational domain was a square grid $14.7 \,\si{km} \times 14.7 \,\si{km}$ in size, and was evolved using four different resolutions with uniform grid spacing: $50\times50$, $100\times100$, and $200\times200$ grid points. For this test, we evolve using the Cowling approximation, meaning the metric is held fixed.

In Fig. \ref{fig:tov-star-convergence} we plot the percent error in the maximum density of the star over time for each resolution, rescaled . We see an initial spike in density at the center of the star.  This is caused by relaxation of the surface of the star, creating a disturbance that moves inward.  Density is continuous but not smooth at the surface, so this feature exhibits approximately first-order convergence. After this initial peak settles, we see second-order convergence.  Rest mass error converges away with resolution and is conserved to $\Delta M_0/M_0\approx 5\times 10^{-6}$ for the highest resolution.

\subsection{Differentially Rotating Star}
\label{sec:diff-star}

\begin{figure}
\includegraphics[width=\textwidth]{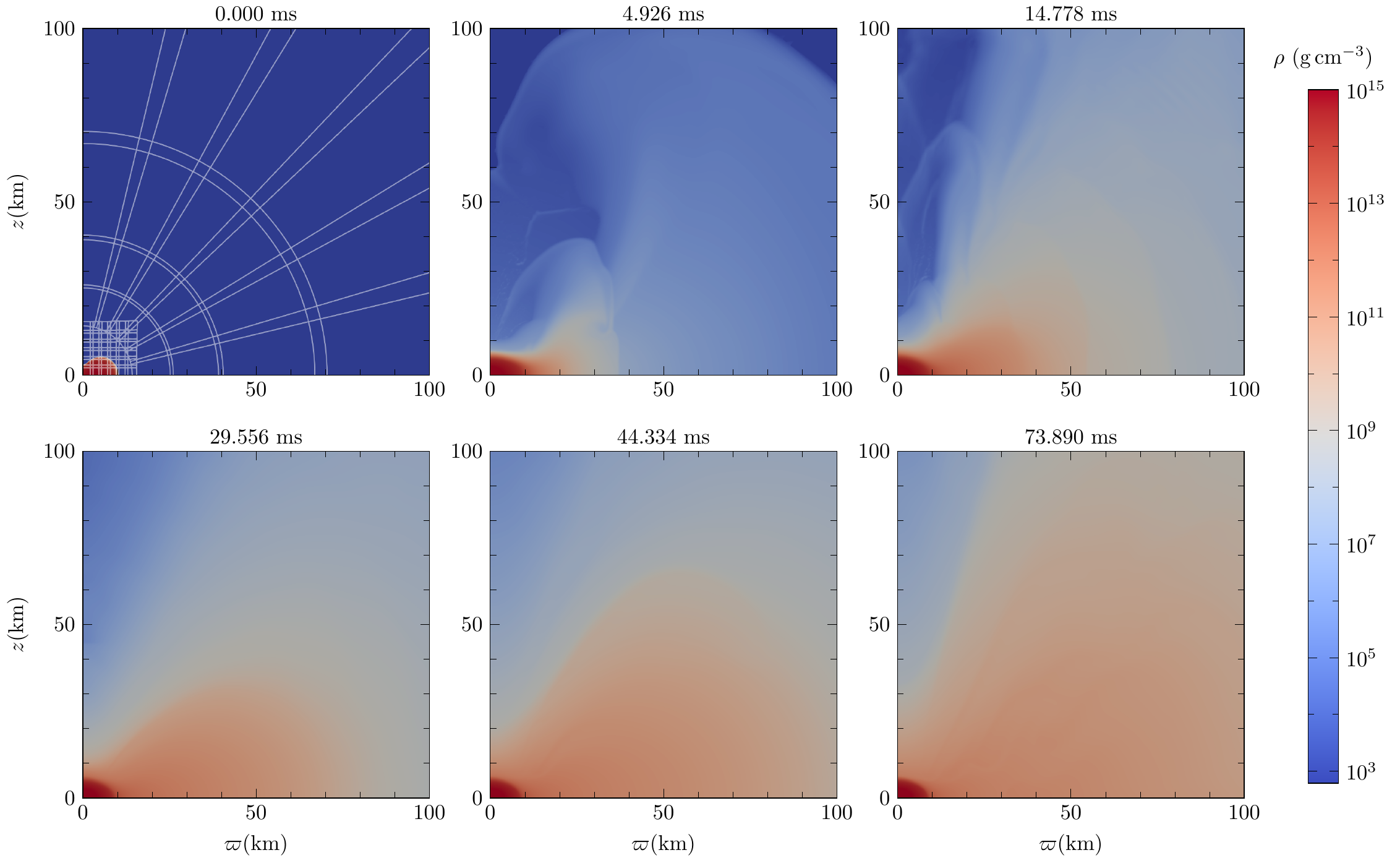}
\caption{Evolution of density profiles of the differentially rotating viscous star with mixing length $\lmix$ = 147 \si{m}. The first panel shows the outlines of each of the overlapping subdomain patches used to construct the computational domain.}
\label{fig:visc-star-time-series}
\end{figure}

\begin{figure}[ht]
\centering
\includegraphics[width=0.75\textwidth]{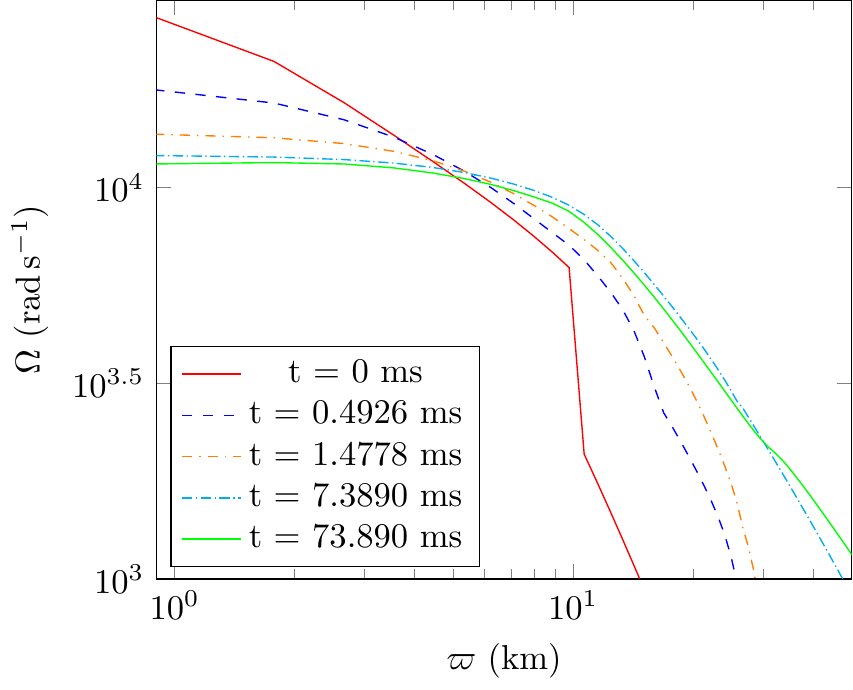}
\caption{Rotational velocity profile of the viscous differentially rotating star in the equatorial plane at multiple times. We see the rotation profile begin to flatten as viscous effects redistribute angular momentum inside the star.}
\label{fig:visc-star-omega-profile}
\end{figure}

\begin{figure}[ht]
\centering
\includegraphics[width=\textwidth]{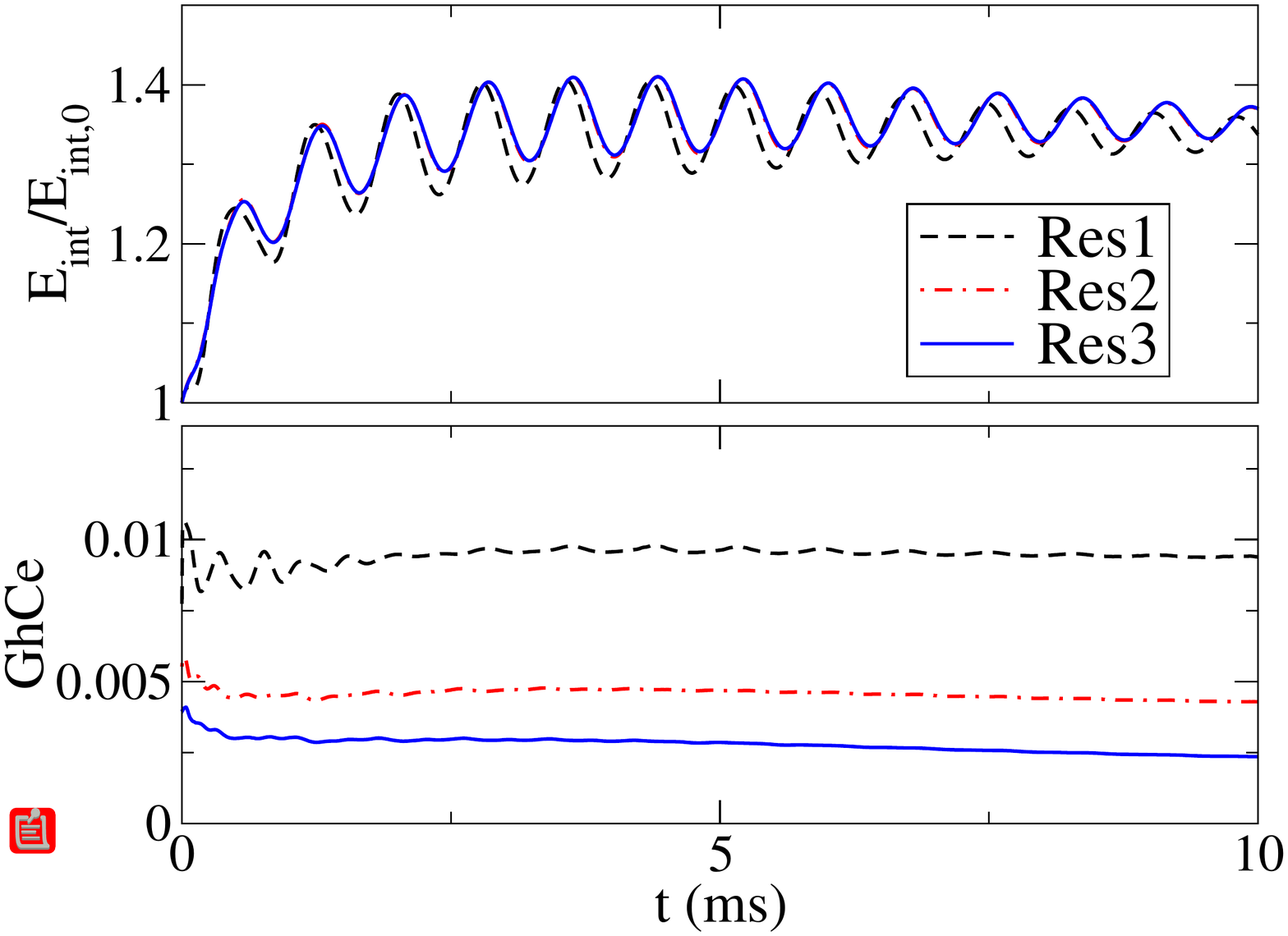}
\caption{Global quantities for three resolutions during the first 10\,ms, the most dynamic phase of the differentially rotating star evolution due to effective viscosity.  Each resolution has 30\% more points in each direction on each subdomain patch of the fluid grid and one extra radial colocation point, two extra angular colocation points for each subdomain of the spectral grid.  Top:  the total internal energy normalized to its initial value.  The two highest resolutions nearly coincide.  Overall, the star heats due to viscosity, but adiabatic fluctuations due to oscillations of the star are also visible.  These modes are excited by a combination of imperfection of the numerically-generated initial data and because the dynamical and viscous timescales are only separated by a factor of$\sim 10^2$.  Bottom:  the normalized generalized harmonic constraint violation.
}
\label{fig:visc-star-convergence}
\end{figure}

We choose a star with very similar profiles and global properties as the differentially rotating star used in Shibata {\it et al}~\cite{Shibata2017a} and likewise use this star to test the evolution of a system under the influence of an effective viscosity.  We point out that this system is not only a useful test of the effective viscosity code, but is designed to resemble the outcome of a binary neutron star merger.  Global quantities such as mass, compaction, and spin resemble such a system, although the form of the rotation profile differs from what is found by merger simulations~\cite{Kastaun:2014fna,Endrizzi2016,Kastaun2016,Ciolfi2017,Hanauske2017,Endrizzi2018,Kiuchi2018,Ciolfi2019}.  Thus, the simulations in~\cite{Shibata2017a} indicate that the interior of the post-merger remnant approaches uniform rotation on a timescale of milliseconds, with the outer layers expanding to form a torus around the central star.  Over the next $\sim 10^2$\,ms, viscous effects acting on the outer star and torus drive an outflow of $\sim 10^{-2}M_{\odot}$ (for sufficiently strong viscosity).  Below, we demonstrate stable hydrodynamic evolution for 100ms, and we confirm the formation of the envelope and massive torus structure able to give rise to outflows using an independent code and different viscosity treatment than~\cite{Shibata2017a}.

The star has an initial baryonic rest mass of 2.64 $\Msun$ and an equatorial radius $R_e = 10.2$ km. We use a piecewise polytropic equation of state, in two pieces, of the form
\begin{equation}
\label{eq:diff-star-eos}
P =
	\left\{\begin{array}{ll}
		\kappa_1 \rho^{\Gamma_1},& \rho \leq \rho_t \\
		\kappa_2 \rho^{\Gamma_2},& \rho \geq \rho_t,
	\end{array}\right.
\end{equation}
where $\kappa_1$ and $\kappa_2$ are polytropic constants, $\Gamma_1$ and $\Gamma_1$ are the polytropic indices, and $\rho_t$ is the density at which we transition between the two pieces. For this star, we choose the polytropic indices to be $\Gamma_1 = 4/3$ and $\Gamma_2 = 11/4$; we set the transition density between the two to be $\rho_t = 1.91 \times 10^{14}$ \si{g.cm^{-3}} and the low-density polytropic constant to $\kappa_1=0.15 {\rm G} \Msun{}^{2/3}$.

The initial rotation profile for the star is given by $u^t u_\phi = \hat{A}(\Omega_0 - \Omega)$ where $\Omega_0$ is the angular velocity along the rotation axis and we choose $\hat{A} = 0.8 R_e$.  The initial equilibrium state is supplied by the code of Cook, Shapiro, and Teukolsky~\cite{cook92}.

In order to handle outflows that will occur when viscosity is enabled, we create a computational grid better suited for resolving both the central star and low density outflowing material. Since any outflows that occur will rapidly drop in density and are not expected to have any small detail features of concern after they leave the region of the star, we leverage the utility of the multipatch technique to apply differing grid structures to each zone of interest. In the central region containing the star we employ the same rectangular grid structure as seen in the previous TOV star test, with a resolution of 100$\times$100 grid points. In the outflow region we switch to a polar grid with constant latitude resolution (so that the proper spacing between angularly adjacent points increases with distance from the star).  The polar grid has 50 points in the angular direction (covering $0<\theta<\pi/2$) and 400 points in the radial direction.  We apply a map to the entire grid that allows us to reduce radial resolution at large distances:
\begin{equation}
\label{eq:visc-star-exp-map}
R = r + 2 e^{-\gamma \beta} \sinh(\gamma r),
\end{equation}
where $r$ is the radius in grid coordinates (the coordinates in which radial grid spacing is uniform), and $R$ is the radius in the original quasi-isotropic, asymptotically-Minkowski coordinates. The map provides an approximately linear grid spacing for radii less than $\beta$, which we have chosen to be at 25.85 \si{km}, and then switches to an exponential grid spacing based on $\gamma$, which is chosen such that $r_{\rm outer}=73.5$ km is mapped to $R_{\rm outer}=2205$ km. The pseudospectral grid used for the evolution of Einstein's equations is composed of an inner ball at the center of the star surrounded by a series of spherical shells extending to a distance of 2940 km.  For the gauge choice, we freeze the generalized harmonic gauge function $H_{\alpha}$ to minimize coordinate dynamics.

In this test, we have modified the density floor from our previous implementations to use a floor dependent on radius:
\begin{equation}
\label{eq:visc-star-atm-density-floor}
\rho_0 > \frac{A}{1+R^2} + B,
\end{equation}
where we have chosen $A = 1.62\times10^4$ \si{g.cm^{-3}} and $B = 1.62\times10^{-2}$ \si{g.cm^{-3}}.

For this test, we employ the viscosity treatment described in Section II. To make a comparison with the results of the $\alpha$-viscosity model used in \cite{Shibata2017b}, we devise a mixing length $\lmix$ corresponding to the same kinematic viscosity as a constant $\alpha$.  The $\alpha$-viscosity model is generalized to differentially rotating stars in~\cite{Shibata2017b} by setting 
\begin{equation}
\label{eq:alpha-visc-parameter}
\nu_\alpha = \frac{\alpha c_s^2}{\Omega_e},
\end{equation}
where $c_s$ is again the local sound speed and $\Omega_e$ is the angular velocity of the star at the surface on the equator. Equating this to Eq. \ref{eq:visc-mixing-length}, we can get an approximate relation between the strength of a given mixing length to that of an $\alpha$-viscosity parameter:
\begin{equation}
\label{eq:alpha-mixlen-compare}
\lmix = \frac{\alpha c_s}{\Omega_e}.
\end{equation}
For the current test we set the viscous mixing length to $\lmix$ = 147 \si{m}, giving a comparable viscous strength to $\alpha \approx 0.01$ in the interior of the star.  (For constant $\lmix$, $\alpha$ as defined in Eq.~\ref{eq:alpha-visc-parameter} is not exactly constant.)  The timescale for viscous angular momentum transport is approximately $R^2/\nu$. Using Eq. \ref{eq:visc-mixing-length} gives a timescale on the order of
\begin{equation}
\label{eq:visc-timescale}
t_{\rm visc} \sim 10\,\si{ms}\left(\frac{r}{10\,\si{km}}\right)^2 \left(\frac{\lmix}{147\,\si{m}}\right)^{-1} \left(\frac{c_s}{0.3c}\right)^{-1}.
\end{equation}

As evolution begins the star quickly begins to transport angular momentum outward causing the rotational velocity profile to become flatter. Although the rotation profile does flatten, we see from Fig. \ref{fig:visc-star-omega-profile} that the profile never completely settles into a rigidly rotating state, and retains some differential rotation. This is a feature of this viscosity method~\cite{Duez:2020}.

Qualitatively the outflow near the star produces the expected distribution of material producing a short, low density burst of material as viscosity is enabled, and at later times as more material leaves the star a disk begins to form.  Other material blows farther outward, indicating the beginnings of a viscous-driven outflow noted in~\cite{Shibata2017a} which we do not follow.  The density profiles in Fig.~\ref{fig:visc-star-time-series} are to be compared to Figure 4 in Shibata~{\it et al}~\cite{Shibata2017a}.  We note that even the qualitative agreement we see in the density plots is nontrivial; it requires the correct treatment of the energy equation described in Section~\ref{sec:viscosity}.  Subgrid momentum transport is modeled in~\cite{Shibata2017a} via an Israel-Stewart-type formulation of the relativistic Navier-Stokes equations, which is analytically quite different from our treatment.  Our qualitative agreement with this previous work gives confidence that its results will not prove very sensitive to details of the momentum transport modeling.  As a further check, we supply convergence tests of the spacetime and fluid evolution in Fig.~\ref{fig:visc-star-convergence}.

\subsection{Magnetized Disk}

\begin{figure}
\includegraphics[width=\textwidth]{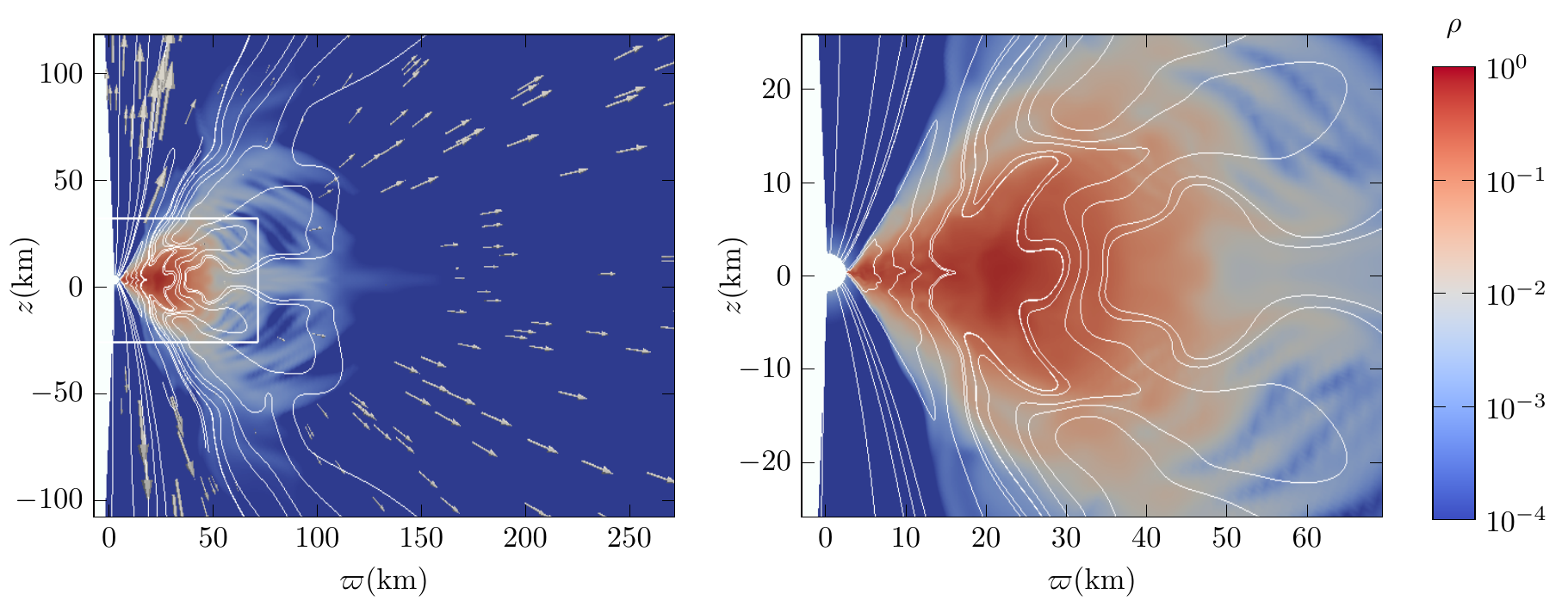}
\caption{Magnetized disk with field lines (for field with toroidal component projected out) at time $t=1500M$, with $M$ the mass of the black hole (which we set equal to one), for evolution on a $256^2$ grid.  Note that the grid extends slightly to the left of the axis because of the symmetry ghost zones.  The right panel shows the region highlighted by the white box in the left panel. Magnetic field and velocity fields are averaged over the time $1000<t/M<1500$.  The initial maximum density in the torus is chosen to be unity.  Profiles are plotted in Kerr-Schild coordinates.  The longest velocity vector arrows close to the poles far from the black hole correspond to speed very close to $1=c$.}
\label{fig:magnetized-disk}
\end{figure}

We evolve a standard axisymmetric MHD test problem:  a magnetized torus around a Kerr black
hole.  The initial conditions for this test are matched to the ``fiducial model'' of McKinney and
Gammie~\cite{mckinney:04}.  A black hole with dimensionless spin $J/M^2=0.938$ is surrounded by a
Fishbone-Moncrief torus~\cite{Fishbone:1976} with inner edge at $r_{\rm BL}=6M$ and specific angular
momentum determined by $u^tu_{\phi}=4.281$.  The torus
has initial maximum density $\rho_0=1$ and a $\Gamma=4/3$ equation of state.  The torus initially
has pressure $P=\kappa\rho^{\Gamma}$, with $\kappa=0.00425$, $\Gamma=4/3$.  A
confined poloidal seed field is introduced via the initial vector potential 1-form
$\widetilde{A}=A_0{\rm max}(\rho_0-1,0)\widetilde{d\phi}$, with $A_0$ chosen to make the
maximum ratio of magnetic to gas pressure be around 0.01. We evolve for 3000$M$ on a
spherical-polar grid with inner radius at $r_{\rm BL}=1.32M$ and maximum radius
at 60$M$.  We evolve with $170^2$, $256^2$, $384^2$, and $512^2$ grids, for which
the initial fraction of the magnetic energy at points where the fastest-growing
MRI mode is resolved by at least 10 points is 0.194, 0.44, 0.66, 0.79, respectively.

The Kerr spacetime is written in in Kerr-Schild coordinates. 
We make the standard change of variables for spherical-polar disk simulations:
\begin{eqnarray}
 r &= \sqrt{x^2 + z^2} = e^{x_1}, \\
 \theta &= \pi x_2 + \frac{1}{2}(1-h)\sin(2\pi x_2).
\end{eqnarray}
Setting a uniform grid in $x_1$, $x_2$ concentrates resolution near the black hole and
on the equator. We set $h=0.5$. Finally, because $r\ne r_{\rm BL}$ we compose with a
final coordinate map to map the coordinate spheres $(x^2+z^2)^{1/2}=C$ to
surfaces of constant Kerr radius $r_{\rm BL}=C$. This allows an excision inner boundary
inside the horizon $r_{\rm BL}=r_+$ that conforms better to the horizon shape.

For this run, we use a position-dependent density floor $\rho_0>10^{-5}r^{-3/2} (\kappa/0.00425)^{-3}c^6$.  We also increase $\rho_0$ and $P$ in the
magnetically-dominated region as needed to maintain $b^2/\rho_0<10$ and
$b^2/P<500$, which significantly improves the step size chosen by the adaptive timestepper.  We evolve both with
hyperbolic divergence cleaning and vector potential evolution.  For the
vector potential evolution, we use the generalized Lorentz gauge~\cite{Farris:2012ux}. Simpler gauges, such as the algebraic $\partial_t\widetilde{A}=\vec{v}\cdot\widetilde{B}$
and advective $\partial_t\widetilde{A}=-\mathcal{L}_v\widetilde{A}$ give the same
evolution of gauge-invariant quantities but, after a while, at a drastically reduced timestep,
presumably because the vector potential does not remain as smooth.

The vector potential
evolution benefits from added explicit dissipation.  We apply Kreiss-Oliger
dissipation~\cite{Kreiss:1973aa} to the evolution of $A_i$ and $\Phi$ with a coefficient of 0.001.
(Our dissipation operator is defined as a sum of fourth derivatives with respect to local
coordinates but applied to global components of the relevant evolved variables.) 
Without dissipation, grid-scale ripples appear in the magnetic field atop an otherwise
reasonable field structure. If the coefficient is increased to $10^{-2}$, the main difference
is a slightly lower asymptotic speed in the polar jets. Kreiss-Oliger dissipation is
not needed for divergence cleaning runs; in fact, it destabilizes the magnetic field evolution
near the excision zone. Instead, extra dissipation for divergence cleaning simulations is
obtained by setting the maximum signal speeds in our HLL Riemann solver
for the evolution of $\tilde{B}$ and $\Psi$ to the null speeds.

The qualitative expectations for this problem are well-known and are reproduced for our
runs for both types of B field evolution. Magnetic winding generates
a toroidal magnetic field, while the magnetorotational instability triggers turbulence
in the disk. Matter falls into the black hole at an average rate of about
$\dot{M}\approx 10^{-1}c^3/G$. The poles become magnetically dominated. An outgoing Poynting
flux can be found in this region, and gas accelerates to near the speed of light on the
poles away from the black hole.  In parts of the polar jet region, $b^2/P$ and the Lorentz factor
grow to the limits imposed by our atmosphere algorithm (500 and 20, respectively). 
The magnetic field energy grows for the first 1000$M$, then
saturates, then begins to die away at a steady rate. This decrease of the magnetic field
is not physical but it is expected in any axisymmetric simulation (at least one not enhanced
by dynamo-modeling additions to the induction equation~\cite{Sadowski:2014awa}) because of the anti-dynamo
theorem.  Outside the region close to the poles, a mildly relativistic wind is seen.  The
configuration of the system at $t=1500M$ is shown in figure~\ref{fig:magnetized-disk}.  All
resolutions produce essentially this same configuration.  However, because the MRI is resolved
in more of the disk, the effect of increasing resolution is to enhance turbulence, in that eddies
appear more distinctly and at finer scales for higher resolutions and the accretion rate increases. 
Also, while the total magnetic energy is initially resolution-independent (when it is dominated by
magnetic winding and linear MRI growth), it saturates at a higher level at higher resolution.
(The lowest resolution saturates at a factor of two lower energy than the highest resolution; the
other resolutions are, of course, closer to the highest, with saturation energy increasing monotonically with resolution, although because it is a turbulent problem no clear order of convergence can be identified.)

None of this is newsworthy, although it is reassuring to confirm for the first time that
SpEC can produce magnetically-dominated jets when they are expected. For our purposes,
the main value of this test is that we can check, for a complex, astrophysically interesting MHD
problem, that our code produces no unphysical axis artifacts in any quantity we have checked
($\rho_0$, $v^i$, $B^i$, $b^2/P$). Of course, the axis actually is a special region in this
problem, which is clearly seen in the solution, but this can easily be distinguished from
artifacts of the coordinate singularity because the latter, when they appear (as they do not in this case), have grid-spacing width. 
The absence of such glitches is, in fact, a nontrivial accomplishment.
For divergence cleaning evolutions without factoring of the evolution equations, grid-scale axis artifacts in the velocity
are easily seen, although they can be suppressed by using low-order reconstruction (MC2~\cite{MC})
near the axis. For vector potential evolutions without factoring the computation of $\tilde{B}$ from $\tilde{A}$, axis
glitches become so severe that simulations crash shortly after accretion onto the black
hole begins.

Although the results are qualitatively similar, we consider the vector potential method
superior for this problem, at least with our current implementations. In divergence cleaning
methods, $\Psi$ builds up at boundaries, particularly the excision boundary. The amount
tends to grow with time and we fear would eventually endanger the simulation. Because of it,
magnetic energy fluxes are not reliable in the inner layer of points (while in vector potential
evolutions, the inner layer shows no problems). Presumably the solution would be to improve
the treatment of the magnetic variables at boundaries.

\subsection{Neutrino Radiation}
\label{sec:nutest}

\begin{figure}
\centering
\includegraphics[width=0.75\textwidth]{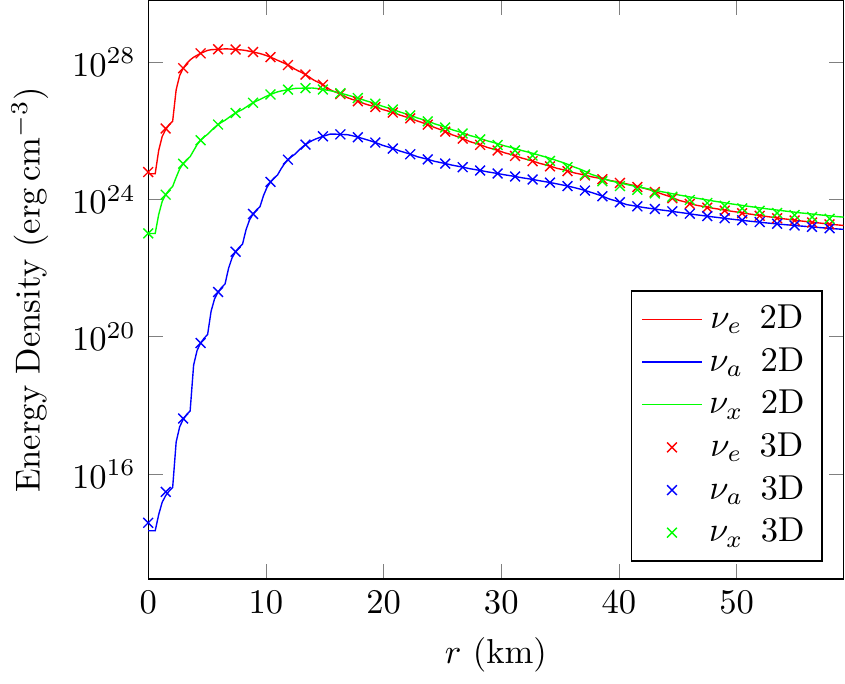}
\caption{Comparison of neutrino energy density along the z-axis for the 2D and 3D spherically symmetric supernova collapse profile at $t$ = 1.5 ms. $\nu_e$, $\nu_a$, and $\nu_x$ represent the electron neutrinos, electron antineutrinos, and heavy lepton neutrinos respectively.}
\label{fig:m1-collapse-energy-comparison}
\end{figure}

\begin{figure}[ht]
\centering
\includegraphics[width=0.75\textwidth]{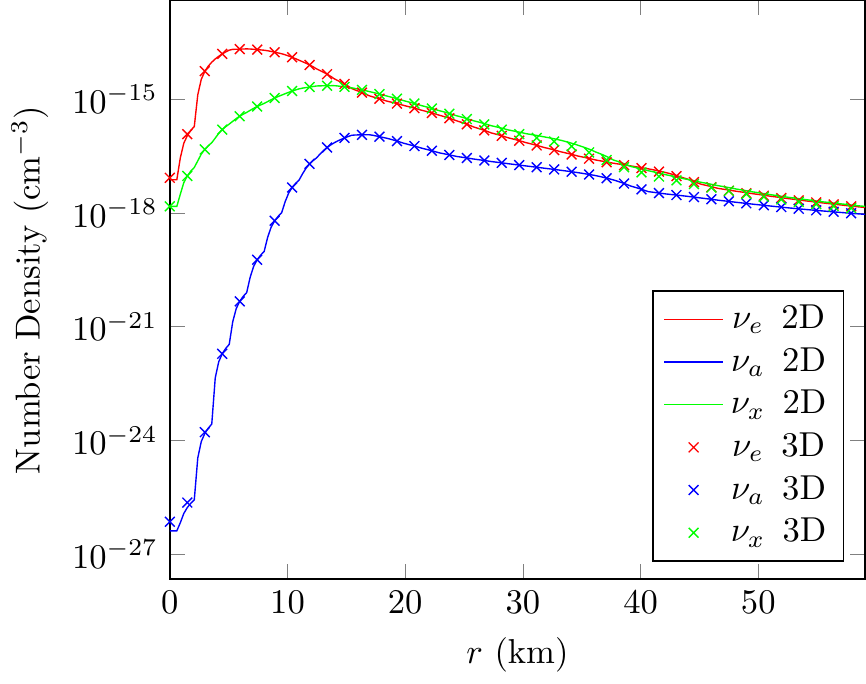}
\caption{Comparison of neutrino number density along the z-axis for the 2D and 3D spherically symmetric supernova collapse profile at $t$ = 1.5 ms. $\nu_e$, $\nu_a$, and $\nu_x$ represent the electron neutrinos, electron antineutrinos, and heavy lepton neutrinos respectively.}
\label{fig:m1-collapse-number-comparison}
\end{figure}

\begin{figure}[ht]
\centering
\includegraphics[width=0.75\textwidth]{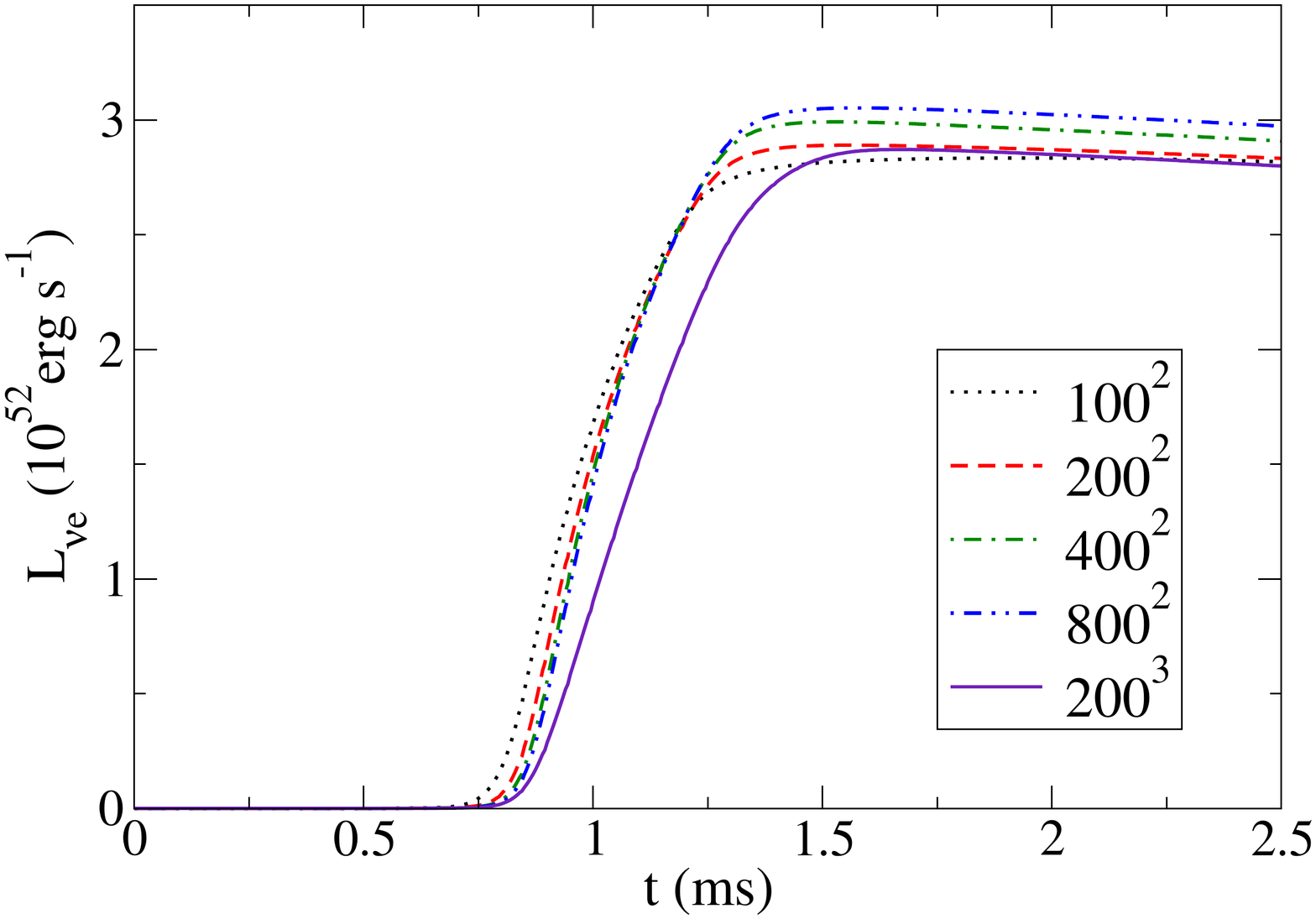}
\caption{Luminosity in electron neutrinos for 3D and 2D evolutions at various resolutions.  The luminosity is calculated by integrating the energy flux through the outer boundary.  Thus, the 3D curve reaches its peak slightly later, since the outer corner of the 3D grid is at $\sqrt{3}\times$300\,km as opposed to the 2D corner at $\sqrt{2}\times$300\,km.}
\label{fig:m1-Lnu}
\end{figure}


Initial testing of the axisymmetric neutrino code was performed by comparing the results obtained from the spherically symmetric post-bounce supernova profile used in \cite{Foucart2016} in both 2D axisymmetry with equatorial symmetry and in 3D using octant symmetry.  The density ranges from 4$\times 10^{14}$g cm${}^{-3}$ at the center of the protoneutron star to $10^{7}$g cm${}^{-3}$ at the outer boundary.  The temperature peaks at 22\,MeV at a radius of 13\,km.  The profile covers a wide range of neutrino opacity from very opaque to very transparent; the neutrinosphere is at a radius of around 30\,km.  In this test we evolve the moments of the neutrino distribution function, fluid temperature, and fluid composition (the electron fraction $Y_e$) for a 1D profile constructed as a spherical average of a 2D core collapse simulation 160 ms after bounce. The velocity of the fluid is set to zero.

We perform this test in 2D on a square grid with length 300 km and a resolution of $200\times200$ grid points. In 3D, we use a cube with the same length of 300 km and a resolution of $200\times200\times200$ grid points. Both systems were evolved for 1.5 ms using a fixed timestep of $4.9\times10^{-3}$ ms to ensure that no error was introduced from possible differences between 2D and 3D in the adaptive timestepper. The 2D test used 24 processing cores and required 2.41 core-hours of run time, whereas the 3D test on 48 cores required 470.27 core-hours, achieving a speed up factor of $\sim$195. We see very strong agreement in results between the 2D and 3D results, as seen in Fig.~\ref{fig:m1-collapse-energy-comparison} and \ref{fig:m1-collapse-number-comparison}.  We also see acceptable agreement in the neutrino luminosity; for the electron neutrinos (the neutrino flavor whose luminosity shows a clearest settled value), both 2D and 3D settle to within about a percent of each other at $L_{\nu_e}=3\times 10^{52}$ erg\,s${}^{-1}$.  By carrying out 2D simulations at other resolutions, we confirm that the 2D vs. 3D agreement is within the truncation error, as shown in Fig.~\ref{fig:m1-Lnu}. 
Although $L_{\nu_e}$ is consistent between resolutions up to differences under 10\%, demonstrating a clear order of convergence is difficult.  This is because of the very sharp behavior of the neutrino fluxes near the neutrinosphere, for which one would require extremely high resolution to be in the convergence regime.

The agreement between 2D and 3D might seem trivial since the 2D and 3D grids are closely matched, but the polar transformation significantly
alters the flux divergence and metric derivative source terms considered separately.  Also, factoring is essential for
avoiding strong axis artifacts.

\section{Conclusion} \label{sec:conclusion}
We have implemented an axisymmetric evolution of the general relativistic hydrodynamics equations through modification of the local coordinate transformations of a multipatch scheme. Without the appropriate factoring of singular terms from spatial derivatives near the symmetry axis, we find that unphysical errors grow in evolved quantities. Testing of this method, with factoring of singular terms applied, produces results that compare favorably to full 3D simulations at a fraction of the required computational time. Since only minimal modification of the implementation of the evolution equations in 3D was required, this method provides a path for a quick application of axisymmetric evolution to codes that make use of computational domains with local coordinate transformations.

We plan to move forward using this method in order to study the effects of a wide variety of physical parameters on binary post-merger environments that require evolution on secular timescales that we have been unable to explore in the past. Additionally, although our method currently evolves Einstein's equations in 3D using spectral methods, we would also like to extend axisymmetry to the evolution of those equations as well.

\ack
J.J. would like to acknowledge Guy Worthey and Sukanta Bose for providing useful comments on an earlier draft of this paper. M.D. would like to acknowledge useful discussions with Thomas Baumgarte,
which helped M.D. overcome his prejudice against analytical treatment of
coordinate singularities. 
J.J. gratefully acknowledges
support from the Washington NASA Space Grant Consortium, NASA Grant NNX15AJ98H.
M.D gratefully acknowledges
support from the NSF through grant PHY-1806207.
F.F. and A.K. gratefully acknowledge
support from the NSF through grant PHY-1806278, from NASA through grant 80NSSC18K0565, and DOE-CAREER grant DE-SC0020435.
H.P. gratefully acknowledges support from the
NSERC Canada. L.K. acknowledges support from NSF grant
PHY-1606654 and PHY-1912081. F.H. and M.S. acknowledge support from NSF Grants
PHY-170212 and PHY-1708213. F.H., L.K. and M.S. also thank
the Sherman Fairchild Foundation for their support.

\bibliographystyle{iopart-num}
\bibliography{axisym-code-paper}

\end{document}